\documentclass[a4paper,fleqn,usenatbib]{mnras}

\usepackage{lscape}

\usepackage[T1]{fontenc}
\usepackage{ae,aecompl}
\usepackage{comment}

\usepackage{graphicx}	
\usepackage{epstopdf}
\usepackage{rotating}
\usepackage{longtable}
\usepackage{multirow}
\usepackage{array}
\usepackage{gensymb}
\usepackage{xcolor}
\usepackage{csquotes}
\usepackage{ragged2e}
\newcommand{\me}{\mathrm{e}^}

\newcommand{\C}{$\epsilon 6$}
\newcommand{\D}{$\epsilon 8$}

\newcommand{\fifty}{50-$\delta 30$-$f 0.25$-$\epsilon 2.5\,$}
\newcommand{\hunA}{100-$\delta 20$-$f 0.25$-$\epsilon 3.5\,$}
\newcommand{\hunB}{100-$\delta 20$-$f 0.3$-$\epsilon 5$\,}
\newcommand{\hunC}{100-$\delta 20$-$f 0.25$-$\epsilon 6$\,}
\newcommand{\hunD}{100-$\delta 20$-$f 0.3$-$\epsilon 8$ \,}
\newcommand{\tenA}{10-$\delta 200$-$f 0.2$-$\epsilon 3.5$ \,}
\newcommand{\tenB}{10-$\delta 200$-$f 0.25$-$\epsilon 3.5$-$\eta 15$ \,}
\newcommand{\simA}{25-$\delta 30$-$f 0.1$-$\epsilon 4 \,$}
\newcommand{\simB}{25-$\delta 30$-$f 0.14$-$\epsilon 4\,$}
\newcommand{\simC}{25-$\delta 30$-$f 0.2$-$\epsilon 4\,$}
\newcommand{\simD}{25-$\delta 30$-$f 0.2$-$\epsilon 2\,$}
\newcommand{\simE}{25-$\delta 30$-$f 0.2$-$\epsilon 2$-$\eta 15\,$}
\newcommand{\simF}{25-$\delta 30$-$f 0.25$-$\epsilon 2.5\,$}
\newcommand{\simG}{25-$\delta 50$-$f 0.3$-$\epsilon 3\,$}
\newcommand{\simH}{25-$\delta 50$-$f 0.5$-$\epsilon 1\,$}

\def\lesssim{\mathrel{\hbox{\rlap{\hbox{\lower4pt\hbox{$\sim$}}}\hbox{$<$}}}}
\def\gtrsim{\mathrel{\hbox{\rlap{\hbox{\lower4pt\hbox{$\sim$}}}\hbox{$>$}}}}

\title[tSZ effect from the EoR]{The Thermal Sunyaev-Zel'dovich Effect from the Epoch of Reionization}

\author[I.~T.~Iliev et al.]{Ilian T. Iliev$^{1}$\thanks{E-mail: I.T.Iliev@sussex.ac.uk},
  Azizah R. Hosein$^{1}$, 
  Jens~Chluba$^2$, Luke~Conaboy$^{3,1}$, David Attard$^1$, \newauthor Rajesh~Mondal$^4$, Kyungjin~Ahn$^5$, 
  Stefan~Gottl\"ober$^6$, Joseph~Lewis$^{7,8}$, Pierre~Ocvirk$^9$, \newauthor Hyunbae~Park$^{10}$, Paul~R.~Shapiro$^{11}$, Jenny~G.~Sorce$^{12,13,6}$, Gustavo Yepes$^{14,15}$\\
$^{1}$ Department of Physics \& Astronomy, University of Sussex,
  Brighton, BN1 9QH, UK\\
$^{2}$ Jodrell Bank Centre for Astrophysics, School of Physics and Astronomy, The University of Manchester, Oxford Road, Manchester, M13 9PL, UK\\
$^{3}$ School of Physics and Astronomy, University of Nottingham, University Park, Nottingham, NG7 2RD, UK\\
$^{4}$ Department of Physics, National Institute of Technology Calicut, Calicut 673601, Kerala, India\\
$^{5}$ Chosun University, 375 Seosuk-dong, Dong-gu, Gwangjiu 501-759, Korea\\
$^{6}$ Leibniz-Institut für Astrophysik Potsdam (AIP), An der Sternwarte 16, D-14482 Potsdam, Germany\\
$^{7}$ Zentrum für Astronomie der Universität Heidelberg, Institut für Theoretische Astrophysik, Albert-Ueberle-Straße 2, D-69120 Heidelberg, Germany \\
$^8$ Max-Planck-Institut für Astronomie, Königstuhl 17, D-69117 Heidelberg, Germany\\
$^{9}$ Université de Strasbourg, Observatoire astronomique de Strasbourg, UMR 7550, F-67000 Strasbourg, France\\
$^{10}$Lawrence Berkeley National Laboratory, CA 94720-8139, USA; Berkeley Center for Cosmological Physics, UC Berkeley, CA 94720, USA\\
$^{11}$ Department of Astronomy and Texas Center for Cosmology and Astroparticle Physics, The University of Texas at Austin, Austin, TX 78712-1083, USA\\
$^{12}$ Univ. Lille, CNRS, Centrale Lille, UMR 9189 CRIStAL, F-59000 Lille, France\\
$^{13}$ Universit\'e Paris-Saclay, CNRS, Institut d'Astrophysique Spatiale, 91405, Orsay, France\\
$^{14}$ Departamento de F\'{\i}sica Te\'orica, Modulo 8, Facultad de Ciencias, Universidad Aut\'onoma de Madrid, 28049 Madrid, Spain\\
$^{15}$ CIAFF, Facultad de Ciencias, Universidad Aut\'onoma de Madrid, 28049 Madrid, Spain}

\date{Accepted XXX. Received YYY; in original form ZZZ}

\pubyear{2024}

\begin{document}
\maketitle

\begin{abstract}
The thermal Sunyaev-Zel'dovich (tSZ) effect arises from inverse Compton scattering of low energy
photons onto thermal electrons, proportional to the integrated electron pressure, and is usually
observed from galaxy clusters. However, we can expect that the Epoch of Reionization (EoR) also contributes
to this signal, but that contribution has not been previously evaluated. In this work we analyse a suite
of fully-coupled radiation-hydrodynamics simulations based on RAMSES-CUDATON to calculate and study
the tSZ signal from the Reionization Epoch. We construct lightcones of the electron pressure in the
intergalactic medium for $6 \lesssim z$ to calculate the resulting Compton y-parameters. We vary the
box sizes, resolutions and star formation parameters to investigate how these factors affect the tSZ
effect. We produce plots of maps and distributions of $y$, as well as angular temperature power spectra
of the tSZ signal obtained from integrating the lightcones constructed for each simulation. We find that the tSZ signal from reionization is generally sub-dominant to the post-reionization one at larger scales ($\ell<10^4$), but can contribute non-trivially and potentially contaminate the measured signals. At scales probed by current
experiments like SPT ($\ell\sim10^3-10^4$), we find that the tSZ signal power spectrum from reionization
contributes at roughly a percent level compared to the current templates,  with the quadratic Doppler effect contributing an additional $\sim10\%$ to the tSZ signal. At smaller scales the tSZ from reionization peaks and can potentially dominate the total signal
and is thus a potentially much more important contribution to take into account in any future, more sensitive experiments.

\end{abstract}

\begin{keywords}
cosmic background radiation -- cosmology: theory -- dark ages, reionization, first stars -- intergalactic medium -- large-scale structure of Universe -- radiative transfer 
\end{keywords}



\section{Introduction}\label{sec:intro}

The epoch of reionization (EoR) marks the last global phase transition of the
intergalactic medium, from cold and largely neutral state to the hot and fully
ionized one that we observe today. It is the period in cosmic history during
which the first luminous objects formed, driving this process. This process
lasted for up to a billion years and was likely very patchy, with some volumes
around the first sources ionized very early, and the deep voids far from any
sources finishing this much later.

Although ongoing and upcoming surveys (e.g. with LOFAR, SKA, and JWST) will
give us considerably more insight into this epoch, leading into its first
direct detection, much still remains unknown. 
Low-frequency radio surveys will allow us to obtain topographical images of
the neutral hydrogen gas from the emitted 21-cm signal, while infrared imaging
will shed more light on bright luminous sources at much higher redshifts than
are currently available. Additionally, the EoR imprints information into the
CMB. While the primary anisotropies of the CMB store information about the
primordial density fluctuations at last scattering, the secondary anisotropies
are due to interactions between the CMB photons and matter along their paths
to us. Therefore, the latter can shed light on the structures and physics in
the universe since the last scattering. One of the contributors to these
secondary CMB anisotropies is the Sunyaev-Zel'dovich (SZ) effect.

The SZ effect \citep{Sunyaev:1972,Sunyaev:1980a,Sunyaev:1980b} \citep[see also
  reviews by ][]{Rephaeli:1995,Birkinshaw:1999, Carlstrom:2002, Kitayama:2014, Mroczkowski:2019} arises from the inverse Compton
scattering of CMB photons by energetic electrons. This transfers energy to
the photons and could distort the CMB spectrum from that of a pure black-body.
There are two types of SZ effect: thermal (tSZ) and kinetic (kSZ). The tSZ
effect is caused by thermally-moving electrons, while the kSZ effect arises
from large-scale bulk motions of electrons. The tSZ effect changes both the
energy and the spectrum of the radiation, while kSZ leaves the spectrum as
black body, but shifted to higher energies. An example of both these effects
can be seen in galaxy clusters - when photons travel through a cluster, they
are scattered by the electrons in the hot intra-cluster gas
($T\sim10^{7} - 10^{8}$ K) which are moving both thermally (tSZ) and with the
overall bulk motion of the cluster itself (kSZ).


The kSZ effect has been proposed as a probe of the EoR, both on its own 
\citep[see e.g.][]{Iliev:2007,Iliev:2008,Mesinger:2012,Park:2013}, and in
terms of its cross-correlation with the redshifted 21-cm emission
\citep[e.g.][]{Jelic2010,Tashiro2010}. The reionization
process creates ionized patches, whose bulk motions yield fluctuating kSZ
secondary anisotropies at the scales corresponding to the patch sizes. These
works showed that EoR has appreciable contribution to the small-scale
($\ell>3000$) CMB anisotropies.

In contrast, to date the tSZ effect has largely been used for detection and
studies of galaxy clusters at lower redshift, but has not been not considered
in EoR context. Due to its characteristic spectral signature (the tSZ effect
lowers the intensity of the photons at frequencies $< 218$ GHz, and raises it
at frequencies $> 218$ GHz), the tSZ effect is used as redshift-independent
approach to detect clusters. It is also useful for measuring the thermal energy
of the intercluster gas and measurements of the Hubble constant.
The tSZ effect is also one of the foregrounds which contaminate CMB
measurements. Templates of its angular power spectrum are thus used for the
purpose of cleaning the primordial CMB signal \citep[e.g.][]{Shaw:2010}.

The tSZ effect is a measure of the integrated electron pressure along the line
of sight. During the EoR the galaxy groups and clusters have not yet formed,
and the typical electron gas temperatures caused by photoionization are of
order tens of thousands of K, rather than the million-degree temperatures of
intercluster gas. However, due to the effects of Hubble expansion
($\bar{\rho}\propto(1+z)^3$), the matter density of both the IGM and halos are
much higher than it is later on, resulting in correspondingly higher gas
pressure. We can therefore reasonably expect that the EoR-produced electron
pressure may provide a non-trivial contribution to the total observed effect.
Furthermore, since the tSZ effect from EoR has some contribution from the
diffuse IGM in the ionized patches, the spatial structure of the temperature
fluctuations should be different from the one produced by galaxy clusters,
potentially providing a new interesting probe of cosmic reionization.
The aim of this paper is to investigate the strength of this effect and its
detectability. A significant contribution from this previously neglected
contribution to tSZ would have consequences on cluster measurements and would
require updating models used to create templates for the tSZ angular power
spectrum.

We investigate the tSZ signal arising from the EoR by analysing the
hydrodynamical data produced by a series of fully-coupled
radiation-hydrodynamics simulations ran using the RAMSES-CUDATON code
\citep{Teyssier:2002,Aubert:2008}. Our baseline simulation is Cosmic Dawn II
\citep[CoDa II;][]{Ocvirk:2018}. The results of further six auxiliary
simulations with varying volumes, and resolutions are used to investigate
the effects of these variations and of varying star formation parameters
on the tSZ signal.

The rest of this paper is organised as follows. We first describe the physics
behind the tSZ effect (Sec.~\ref{sec:theory}) and the simulations used in this
project (Sec.~\ref{sec:simulations}). We then outline our methodology
(Sec.~\ref{sec:method}) and present our results (Sec.~\ref{sec:results}) and
end with our conclusions (Sec.~\ref{sec:conculsion}). The data analysis for this 
work was performed in part using the seren3 Python 
package\footnote{\url{https://github.com/sully90/seren3} }.

\section{The Sunyaev-Zel'dovich Effect}
\label{sec:theory}
The SZ effect arises from the inverse Compton scattering of CMB photons by free
electrons. When CMB photons travel through a cloud of such electrons, the
probability that they will scatter is dictated by the Thomson scattering optical
depth,
\begin{equation}
  \tau_\mathrm{e} = \sigma_\mathrm{T} \int n_\mathrm{e}\ \mathrm{d}l \sim 2\times10^{-3}
  \left( \frac{n_\mathrm{e}}{10^{-3}\ \mathrm{cm^{-3}}} \right) \left( \frac{l}{\mathrm{Mpc}}\right),
\end{equation}
where the integral is performed along the line-of-sight, $n_\mathrm{e}$ is the
electron number density, and the fiducial values denoted are typical for galaxy
clusters, where the SZ effect is most commonly observed. Although inverse
Compton scattering occurs in a variety of scenarios, the SZ effect usually
refers to the scattering of CMB photons in the GHz to THz range of frequencies
on non- or mildly relativistic electrons.

\subsection{The Thermal Sunyaev-Zel'dovich Effect}
When CMB photons pass through the hot intra-cluster medium (ICM), there is a
$\sim 1\%$ chance that it will interact with one of the energetic electrons in
the plasma. The scattered photon experiences an energy boost of
$\sim 4 \frac{k_\mathrm{B}T_\mathrm{e}}{m\mathrm{e}c^2}$
\citep[see e.g.][]{Rybicki:1979, Sazonov:2000}, causing a distortion in CMB
intensity given by
\begin{equation}
    \Delta I_\nu \approx I_0 y \frac{x^4 \me{x}}{\left( \me{x}-1\right)^2}
    \left( x \frac{\me{x}+1}{\me{x}-1} -4 \right)
    \equiv I_0 y g(x),
\end{equation}
where $x = \frac{h_\mathrm{P} \nu}{k_\mathrm{B}T_\mathrm{CMB}} \approx \frac{\nu}{56.8\ \mathrm{GHz}}$
is the dimensionless frequency, $y$ is the Compton $y$-parameter (defined
below), and
\begin{equation}
  I_0 = \frac{2\left(k_\mathrm{B}T_\mathrm{CMB}\right)^3}{\left( h_\mathrm{P}c\right)^2}
  = 270.33 \left( \frac{T_\mathrm{CMB}}{2.73\ \mathrm{K}}\right)^3\ \mathrm{MJy\ sr^{-1}}.
\end{equation}
If $\frac{\Delta I_\nu}{I_\nu} \ll 1$, the signal can be expressed in terms of
the CMB temperature, using the derivative of the Planck function with respect
to temperature:
\begin{equation}\label{eq:freqdep}
    \frac{\Delta T_\mathrm{CMB}}{T_\mathrm{CMB}} \approx y \left( x \frac{\me{x}+1}{\me{x}-1} -4 \right) = y f(x),
\end{equation}
where $f(x)$ is the frequency dependence of the tSZ spectrum in terms of
$\Delta T_\mathrm{CMB}$. When there are electrons with energies in the
relativistic regime, a factor $\delta_\mathrm{SZ}(x,T\mathrm{e})$ is
incorporated, and $f(x)$ becomes
\begin{equation}
    f(x) = \left( x \frac{\me{x}+1}{\me{x}-1} -4 \right) \left( 1 + \delta_\mathrm{SZ}\left(x,T\mathrm{e}\right) \right).
\end{equation}
For more detailed interpretations of the relativistic corrections to the tSZ
effect, see e.g. \citet{Challinor:1998, Nozawa:2006, Chluba:2013}. We will
neglect such effects here. In the non-relativistic case and in the
Rayleigh-Jeans, low-energy limit, we have $f(x\ll1) \rightarrow -2$, while at
high frequencies, $f(x\gg1) \rightarrow x-4$.

The spectral signature of the tSZ can be separated from the kSZ effect around
the null of the tSZ effect, and other temperature fluctuations due to its
characteristic frequency dependence above. Hence, multi-frequency measurements
are necessary to distinguish all these effects. Additionally, the tSZ signal
dominates over the kSZ effect roughly by an order of magnitude for clusters.
This is due to the thermal velocity of electrons ($\sim 10^4$ km s$^{-1}$)
being much higher than the bulk velocity ($\lesssim 10^3$ km s$^{-1}$).

We can see from Equation~\ref{eq:freqdep} that the change in the CMB temperature
is proportional to the Compton $y$-parameter, which is thus used to measure the
magnitude of the tSZ signal. It is defined as the line of sight integral of the
electron pressure,
\begin{equation}\label{eq:y-param}
    y \equiv \frac{k_\mathrm{B}}{m_\mathrm{e}c^2} \int T_\mathrm{e}\ \mathrm{d}\tau_\mathrm{e} = 
    \frac{k_\mathrm{B} \sigma_\mathrm{T}}{m_\mathrm{e}c^2} \int n_\mathrm{e} T_\mathrm{e}\ \mathrm{d}l = 
    \frac{\sigma_\mathrm{T}}{m_\mathrm{e}c^2} \int p_\mathrm{e}\ \mathrm{d}l,
\end{equation}
where $T_\mathrm{e}$ is the temperature of the electrons, and $p_\mathrm{e}$ is the
pressure due to the electrons. It is important to note that the Compton
$y$-parameter is redshift-independent. This can also be explained by considering
that $\Delta T_\mathrm{CMB}$ (and $\Delta I_\nu$) is redshifted the same way as
$T_\mathrm{CMB}$ (and $I_\nu$). Hence, the tSZ effect does not experience a loss in
intensity with redshift (dimming), making it a useful tool for measuring
large-scale structure in the universe. The ICM of typical clusters have electron
temperatures of 5 - 10 keV. For massive clusters with central optical depth
$\sim 10^{-2}$ we have $y \sim 10^{-4}$. More generally, the $y$-parameter for clusters is
typically $y \gtrsim 10^{-5}$. In practice, all clusters above certain mass cutoff
in a given area of sky can be detected with the tSZ effect using CMB frequency
maps of the $y$-parameter.

Another useful observable is the angular power spectrum of the tSZ signal.
\citet{Komatsu:2002} found that the tSZ angular power spectrum is
cosmology-dependent, with $D_l \propto \sigma_8^7 \Omega_\mathrm{b}^2 h^2$.
Later \citet{Shaw:2010} used numerical simulations to refine this scaling to
$D_l \propto \sigma_8^{8.3} \Omega_\mathrm{b}^{2.8} h^{1.7}$. This dependence makes the
tSZ power spectrum a powerful tool for constraining these cosmological parameters
\citep[see e.g.][]{Barbosa:1996, Sievers:2013, Crawford:2014, Hill:2014,
  George:2015, Planck-Collaboration:2015b, Horowitz:2017}.
The tSZ power spectrum is also affected by astrophysical processes, such as AGN
feedback, shock heating, radiative cooling, etc. \citet{Komatsu:2002,
  Battaglia:2010, Sehgal:2010, Shaw:2010, Trac:2011} have produced templates for
the (post-reionization) tSZ power spectrum, fitted for WMAP $\Lambda$CDM cosmology,
while probing its dependence on cosmological and astrophysical effects.
While these studies investigate the tSZ signal in the context of galaxy clusters,
this is not the only possible source of such signal. The aim of our current study
is to determine the tSZ effect arising from the EoR, which previously has largely
been neglected.

It is instructive to consider the mean values of the $y$-parameter, $\left<y\right>$
produced by various large-scale cosmological structures. The tightest
constraint on overall $\left<y\right>$ was done by the COBE-FIRAS experiment and found
$\left<y\right> < 1.5\times10^{-5}$ at the 95\% confidence level \citep{Fixsen:1996}.
\citet{Refregier:2000} performed hydrodynamic simulations comprising only
gravitational forces (with no star formation) as well as analytic calculations
with the Press-Schechter (PS) formalism \citep{Press:1974} to compute the tSZ
signal. For $\Lambda$CDM cosmology, they found $\left<y\right> = 1.67\times10^{-6}$
(simulation) and $2.11\times10^{-6}$ (PS), both values being about an order of
magnitude below the COBE-FIRAS upper limit. Their projected maps of the
$y$-parameter showed clusters having $y>10^{-5}$, and groups and filaments having
$y \sim 10^{-7}$- $10^{-5}$. However, they note that projecting a number of
simulation boxes along the line of sight on the sky would cause the filamentary
objects to be averaged out \citep{da-Silva:2001, Seljak:2001}. They also find
that the majority of the tSZ signal arises from low redshifts ($z<2$). Their
angular power spectra showed the tSZ effect having comparable power to the primary
CMB at $l\sim2,000$, while groups and filaments contributed $\sim 50\%$ of the
power at $l=500$ with $\sim50\%$ of that power being produced at $z\lesssim0.1$.

By combining hydrodynamic simulations and analytic models, \citet{Zhang:2004a}
estimated
\begin{equation}
\left<y\right> = 2.6\times10^{-6} \left(\frac{\sigma_8}{0.84}\right)^{4.1-2\Omega_\mathrm{m}} \left( \frac{\Omega_\mathrm{m}}{0.268} \right)^{1.28-0.2\sigma_8}
\end{equation}
for a flat $\Lambda$CDM WMAP cosmology, with the dominant contribution coming
from $z \sim 1$. More recently, \citet{Hill:2015} used analytic calculations
(including an ICM model
\citep{Hill:2014}, relativistic corrections \citep{Arnaud:2005} and a reionisation
model \citep{Battaglia:2013}) to compute the total mean Compton parameter of the
universe. They found $\left<y\right>_\mathrm{ICM} = 1.58\times10^{-6}$,
$\left<y\right>_\mathrm{IGM} = 8.9\times10^{-8}$ and
$\left<y\right>_\mathrm{EoR} = 9.8\times10^{-8}$
for the contributions from the ICM, IGM and EoR, respectively, which gave a total
mean value of $\left<y\right>_\mathrm{total} = 1.77\times10^{-6}$. The signals from
the IGM and EoR are thus expected to be sub-dominant to the ICM signal. However,
none of these results is based on full radiative-hydrodynamics EoR simulations,
and can thus might not be fully accurate. Furthermore, even if this EoR tSZ signal
is sub-dominant, its accurate estimation would be very useful for providing constraints
on galaxy formation models, feedback mechanisms and the thermal history of the
universe. It could also be a possible source of errors for cluster tSZ measurements. 


\subsection{Second Order Doppler Distortions}
\citet{Zeldovich:1972} showed that the spectral distortions produced by thermally
energetic electrons give rise to a spectrum which can be described as a
superposition of blackbodies with a Compton $y$-parameter, $y = \mathcal{O}(v^2)$.
This $y$-parameter depends on the second order of the electron velocities, $v$,
essentially making it the equivalent of a tSZ effect due to the bulk flow.
\citet{Hu:1994a} investigated the significance of this quadratic Doppler
distortion during the EoR. They showed that when this contribution to the
$y$-parameter is included, Equation~\ref{eq:y-param} becomes
\begin{equation}
    y = \frac{\sigma_\mathrm{T}}{m_\mathrm{e}c^2} \int \left( \frac{1}{3} m_\mathrm{e} n_\mathrm{e} \left<v^2\right> + p_\mathrm{e}
    \right)\ \mathrm{d}l.
    \label{eq:yso}
\end{equation}
They also suggest that the bulk flow contribution would be dominated by the tSZ
effect. According to their estimations, for a CDM cosmology with $\Omega_0 = 1$, the quadratic Doppler effect does not yield a measurable average
(isotropic) Compton-y parameter. 
Below we evaluate this second order Doppler contribution based on our simulations for the currently-favoured cosmological model.

\section{Simulations}\label{sec:simulations}


\begin{table*}
    \centering
    \label{tab:parameters}
    \caption{Summary of the parameters used in the simulations. A few parameters were not varied between our
      simulations: massive star lifetime $t_\star=10$~Myr; supernova energy $E_\mathrm{SN}=10^{51}$~erg;  
      effective photon energy of 20.28 eV;
      effective HI cross-section $\sigma_\mathrm{E}=2.493\times 10^{-22}$ m$^2$;
      and (full) speed of light $c=299 792 458$~m/s.
    } 
    \begin{tabular}{l|c|c|c|c|c}
    \hline
  Simulation set  & CoDa II & 100$\,h^{-1}$~Mpc & 50$\,h^{-1}$~Mpc & 25$\,h^{-1}$~Mpc & 10$\,h^{-1}$~Mpc \\
    \hline
    \multicolumn{6}{|c|}{Setup} \\
    \hline
    Number of nodes (GPUs) & 16384 & 128 & 128 & 32 & 128 \\
    Grid size & $4096^3$ & $1024^3$ & $1024^3$ & $512^3$ & $1024^3$  \\
    Comoving box size $L_\mathrm{box}$ & $64 h^{-1}$ Mpc & $100 h^{-1}$ Mpc & $50 h^{-1}$ Mpc & $25 h^{-1}$ Mpc & $10 h^{-1}$ Mpc \\
    Comoving force resolution d$x$ & 23.1 kpc & 145.5 kpc & 72.8 kpc & 72.8 kpc & 14.6 kpc \\
    Physical force resolution at $z=6$ & 3.3 kpc & 20.8 kpc & 10.4 kpc &10.4 kpc & 2.1 kpc \\
    DM particle number $N_\mathrm{DM}$ & $4096^3$ & $1024^3$ & $1024^3$ & $512^3$ & $1024^3$ \\
    DM particle mass $M_\mathrm{DM}$ & $4.07 \times 10^5 \mathrm{M_\odot}$ & $1.05 \times 10^8 \mathrm{M_\odot}$ & $1.31 \times 10^7 \mathrm{M_\odot}$ & $1.31 \times 10^7 \mathrm{M_\odot}$ &$1.05 \times 10^5 \mathrm{M_\odot}$ \\
    Initial redshift $z_\mathrm{start}$ & 150 & 80 & 80  & 80 & 80 \\
    End redshift $z_\mathrm{end}$ & 5.8 & 4.0, 0.0, 4.1, 1.0 & 0.0 & 5.0 & 6.7 \\
    \hline
    \multicolumn{6}{|c|}{Star formation} \\
    \hline
    Density threshold $\delta_\star$ & $50 \left<\rho\right>$ & $20 \left<\rho\right>$ & $30 \left<\rho\right>$ & $30 \left<\rho\right>$,$50 \left<\rho\right>$ & $200 \left<\rho\right>$ \\
    Temperature threshold $T_\star$ & \textbf{off} & $2 \times 10^4$ K & $2 \times 10^4$ K & $2 \times 10^4$ K & $2 \times 10^4$ K \\
    Efficiency $\epsilon_\star$ & 0.02 & 0.035, 0.05, 0.06, 0.08 & 0.025 & 0.01, 0.02,0.025,0.03,0.04& 0.035 \\
    \hline
    \multicolumn{6}{|c|}{Supernova feedback} \\
    \hline
    Mass fraction $\eta_\mathrm{SN}$ & 10\% & 10\% & 10\% & 10\%, 15\% & 10\%, 15\% \\
    \hline
    \multicolumn{6}{|c|}{Radiation} \\
    \hline
    Stellar particle escape fraction $f_{\mathrm{esc},\star}$ & 0.42 & 0.25, 0.3, 0.25, 0.3 & 0.25 & 0.1,0.14,0.2,0.25,0.3,0.5& 0.2, 0.25 \\
\hline
    \end{tabular}
\end{table*}

All simulations used in this study were performed using the code RAMSES-CUDATON
\citep{Teyssier:2002,Aubert:2008}, which is a fully-coupled, fixed-grid, hybrid
GPU-CPU code combining N-body dynamics, gasdynamics and (GPU-based) radiative
transfer for simulating large-scale structure and galaxy formation. The N-body
dynamics solve for the velocities and positions of collisionless dark matter
particles. The gas is treated by solving the Euler equations using a second-order
Godunov shock-capturing method \citep{Toro:2009}.
The parameters and setup of our simulations are summarised in Table~\ref{tab:parameters}.

\subsection{Cosmic Dawn II}\label{sec:CoDaII}
The Cosmic Dawn (CoDa) II \citep{Ocvirk:2018} is the largest simulation we use
here, which serves as our fiducial case. It has comoving box of $64\ h^{-1}$ Mpc
on a side, with $4096^3$ grid for the gas and radiation dynamics, and $4096^3$
dark matter particles of mass $4.07 \times 10^{5}$ M$_\odot$, with the gravitational
forces solved on a $4096^3$ grid. The simulation uses the Planck 2014 cosmology
\citep{Planck-Collaboration:2014b}: 
$\Omega_{\Lambda}=0.693$, 
$\Omega_{m}=0.307$, 
$\Omega_{b}=0.045$, 
$H_0=67.77$, 
and power spectrum normalization $\sigma_{8}=0.8288$ and slope $n=0.963$. 
The simulation starts at redshift $z=150$ and ends at $z=5.80$.

For storage efficiency, the data for the full resolution grid was reduced to a
coarser grid of $2048^3$ cells. These full-box, lower-resolution data of all the
gas properties, ionizing flux density, and dark matter density field were kept
for all snapshots. Furthermore, a catalogue of the stellar particles and halo
catalogues are available.
For a more detailed description of CoDa II, we direct the reader to
\citet{Ocvirk:2018}.

\subsection{10, 25, 50, and 100 $h^{-1}$ Mpc simulations}\label{sec:sims}
Our suite of auxiliary simulations use cosmology parameters consistent with the
latest constraints from the Planck survey \citep{Planck-Collaboration:2018}:
$\Omega_{\Lambda}=0.682$, 
$\Omega_{m}=0.318$, 
$\Omega_{b}=0.045$, 
$H_0=67.1$, 
and power spectrum normalization $\sigma_{8}=0.833$ and slope $n=0.9611$. The
15 simulations used here vary in volume, spatial resolution and star formation
parameters (see Table~\ref{tab:parameters}).
There are four runs with sides $100\, h^{-1}$ Mpc (149.0 Mpc), one run with side
$50\, h^{-1}$ Mpc (74.5 Mpc), and two runs with sides $10\,h^{-1}$ Mpc (14.9 Mpc),
all with $1024^3$ cells and $1024^3$ dark matter particles. We have also performed
a further 8 runs with sides $25\,h^{-1}$ Mpc (37.25 Mpc) with $512^3$ cells and
$512^3$ dark matter particles. All simulations adopted initial power spectrum of density fluctuations based on CAMB code\footnote{https://camb.readthedocs.io/en/latest/}, except for the $25\,h^{-1}$ volumes which used the \citet{1999ApJ...511....5E} power spectrum. We do not expect this difference to have a notable effect on our results. 

We label our simulations by listing the main parameters being varied, as follows
Boxsize$-\delta$Num1$-f$Num2$-\epsilon$Num3$(-\eta$Num4), where Boxsize is the
size of the simulation volume in $h^{-1}$~Mpc, Num1 is the value of the density
threshold for star formation used, Num2 is the escape fraction per stellar particle,  
Num3 is the star formation efficiency in percents and Num4 is the supernova feedback
load factor (optional, indicated only if different from the fiducial value of 0.1),
for example \simE.

The main differences between these simulations and CoDa II are the feedback
strength, ionizing photon escape fractions and star formation parameters. For
instance, the cooling floor for the interstellar gas is switched off in CoDa II,
i.e. there is no temperature limit below which the cells need to be for stars to
form. However, the temperature floor is turned on in the auxiliary simulations.
The density threshold for star formation was varied between 200 above mean
for the $10\,\rm Mpc\,h^{-1}$ boxes to $20$ for the $100\,\rm Mpc\,h^{-1}$
ones, in order to account for the different grid resolution of these volumes.
With the exception of \tenB, all simulations run from $z = 80$ to the end of
the EoR or later. Simulation \tenB \, was stopped early since its ionization
fraction was evolving much faster than expected and reionization was due to
be completed too early and sooner than observations suggest (see
Fig.~\ref{fig:calib_plots}).

\subsection{Calibration of the simulations}

\begin{figure*}
        \includegraphics[width=0.49\linewidth]{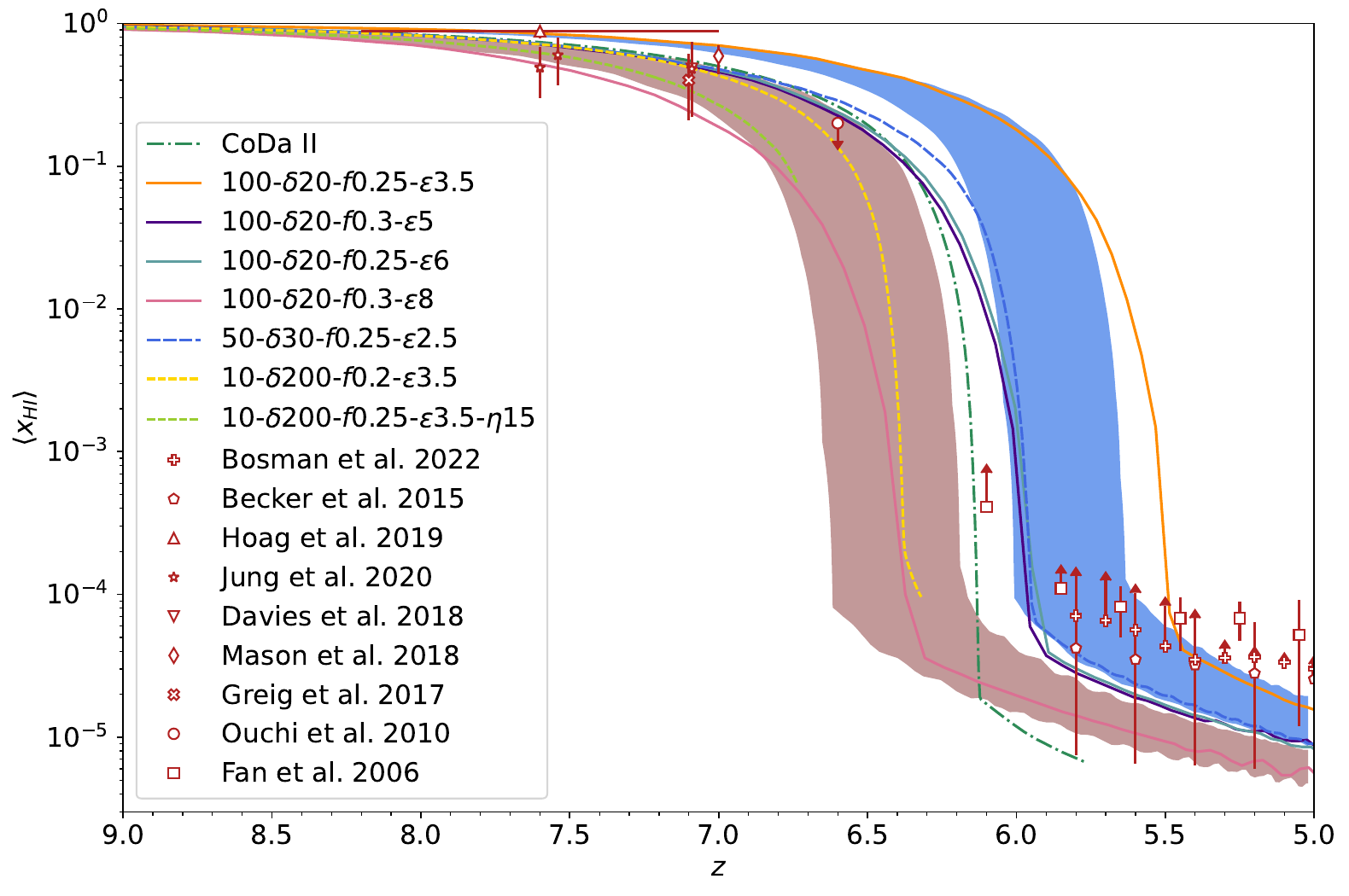}
	\includegraphics[width=0.49\linewidth]{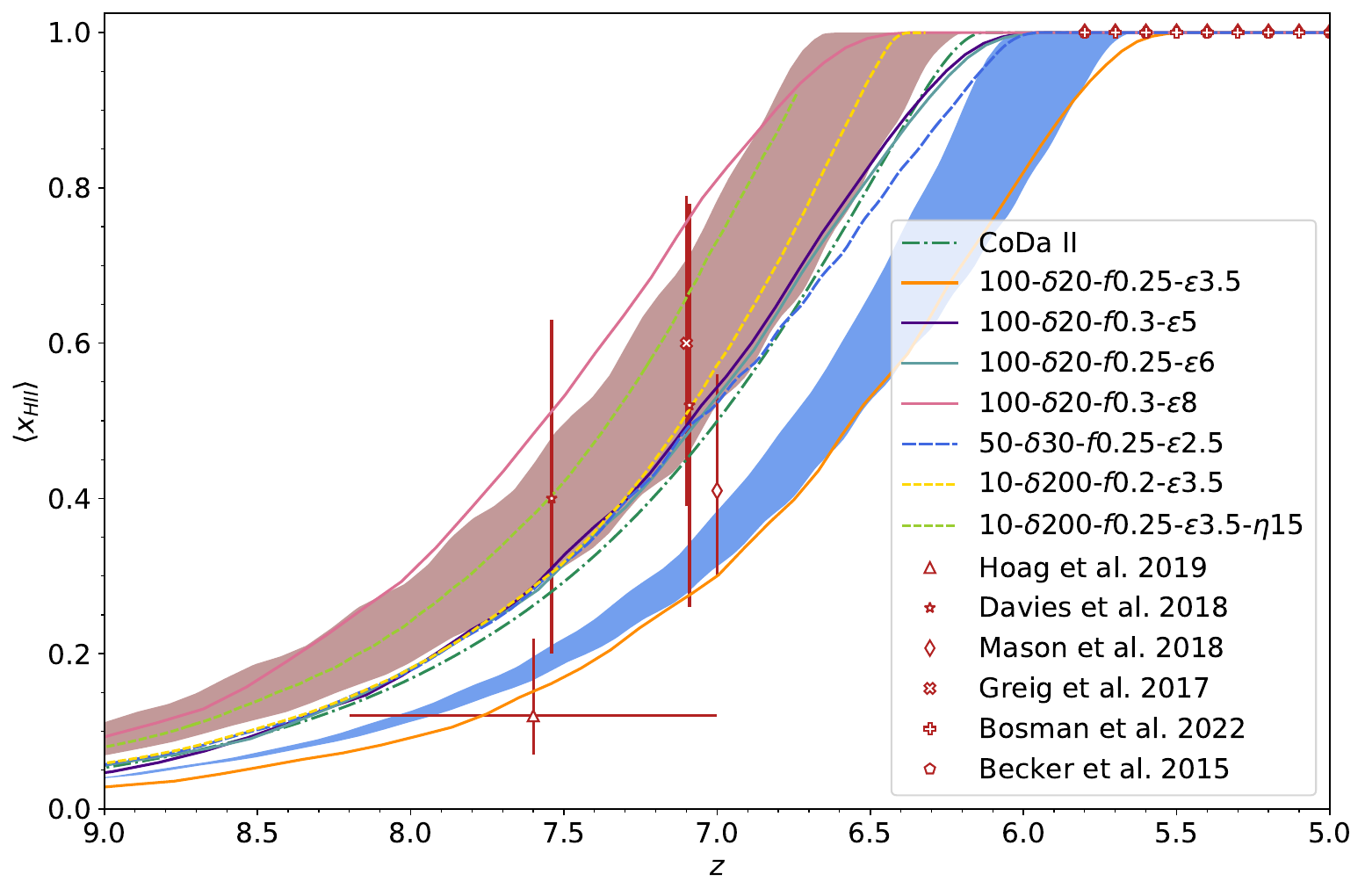}
	\includegraphics[width=0.49\linewidth]{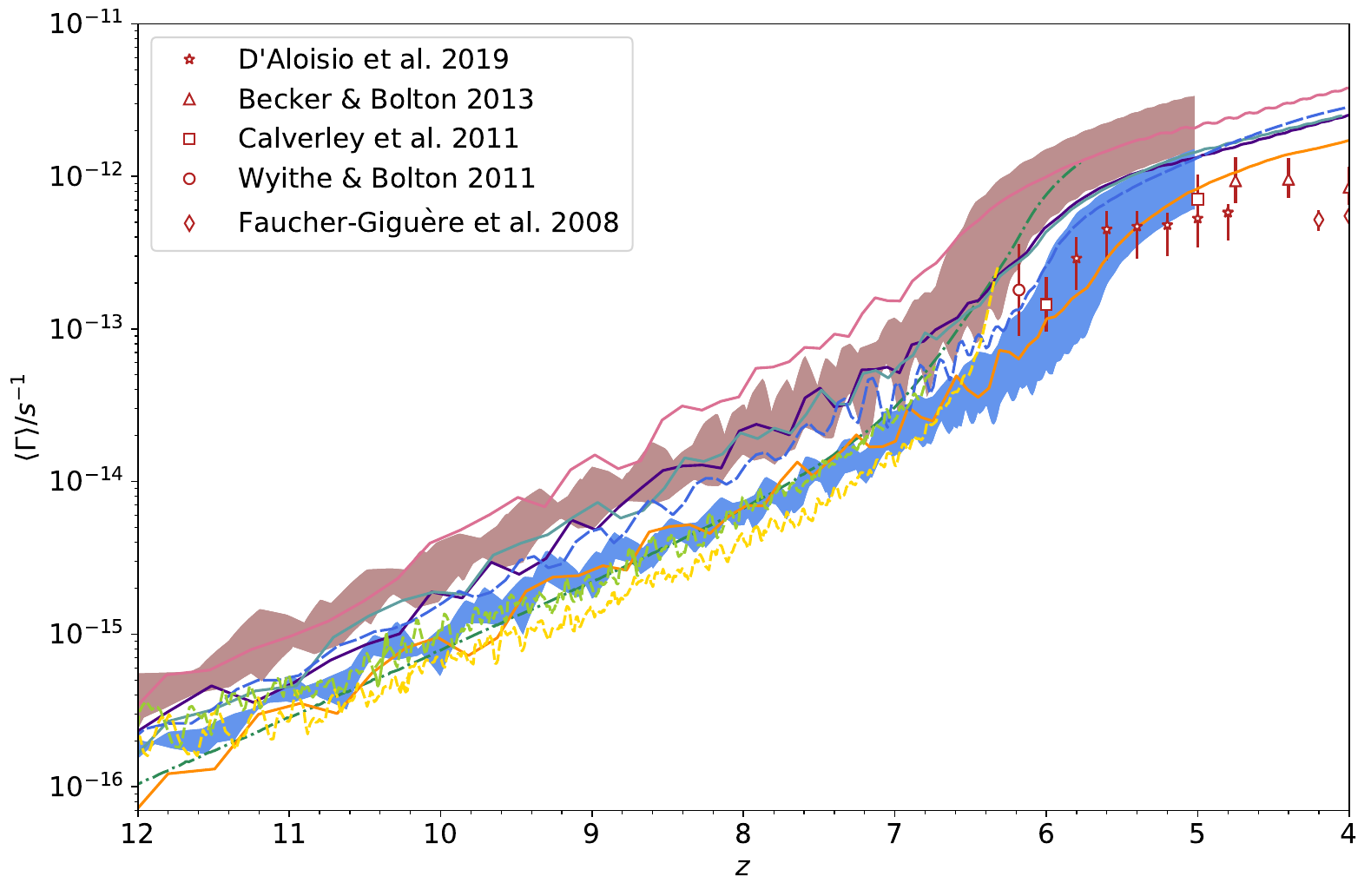}
	\includegraphics[width=0.49\linewidth]{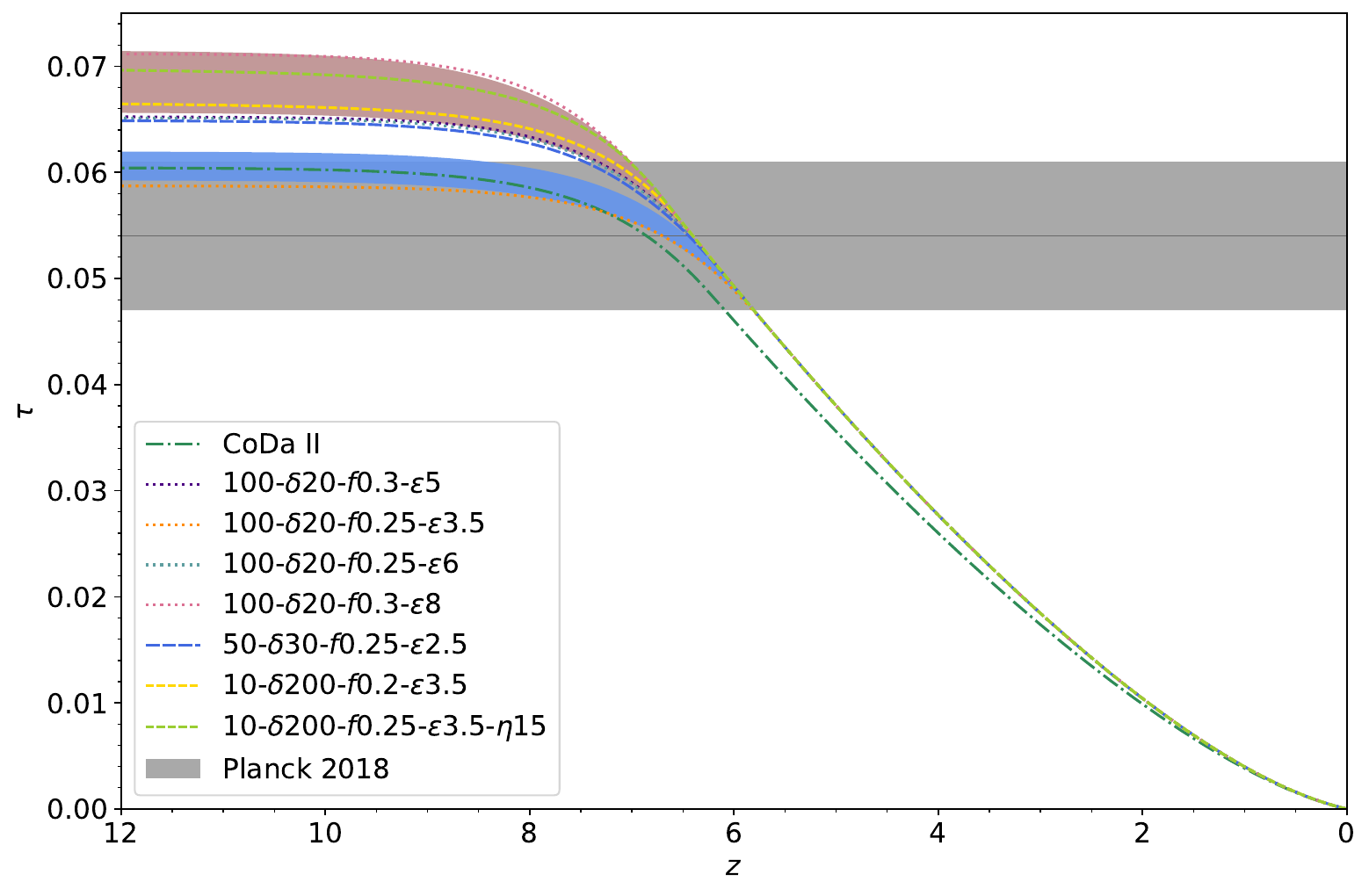}
	\caption{Calibration of our simulations. For clarity not all simulations are shown as
          separate lines, but instead the $25\,h^{-1}$Mpc box models are shown as bands, the
          rose-brown band indicates the range for early-reionization cases (\simB, \simC, \simF
          and \simG), 
          while the blue band includes the late-reionization ones (\simA, \simD, \simE,
          and \simH). The respective observational data points are shown as symbols, as indicated
          on each panel. We show the evolution of the following globally-averaged quantities:
          (top left) Neutral hydrogen fraction
          \citep{Hoag_2019,Davies_2018,Mason_2018,Greig_2017,Ouchi:2010,Fan:2006}
%
          (top right) Ionised hydrogen fraction \citep{Hoag_2019,Davies_2018,Mason_2018,Greig_2017}.
          (middle left) Volume-weighted photoionisation rate
          \citep{DAloisio:2019,Becker_2013,Wyithe:2011,Calverley:2011,Faucher_Giguere_2008}. 
%
          (middle right) Thomson scattering optical depth due to reionisation.
          The measurement from the Planck Collaboration \citep{Planck-Collaboration:2018}
          including its 1-$\sigma$ error (grey shaded area) is shown for comparison.
          %
          \label{fig:calib_plots}}
\end{figure*}

\begin{figure}
  \includegraphics[width=0.99\linewidth]{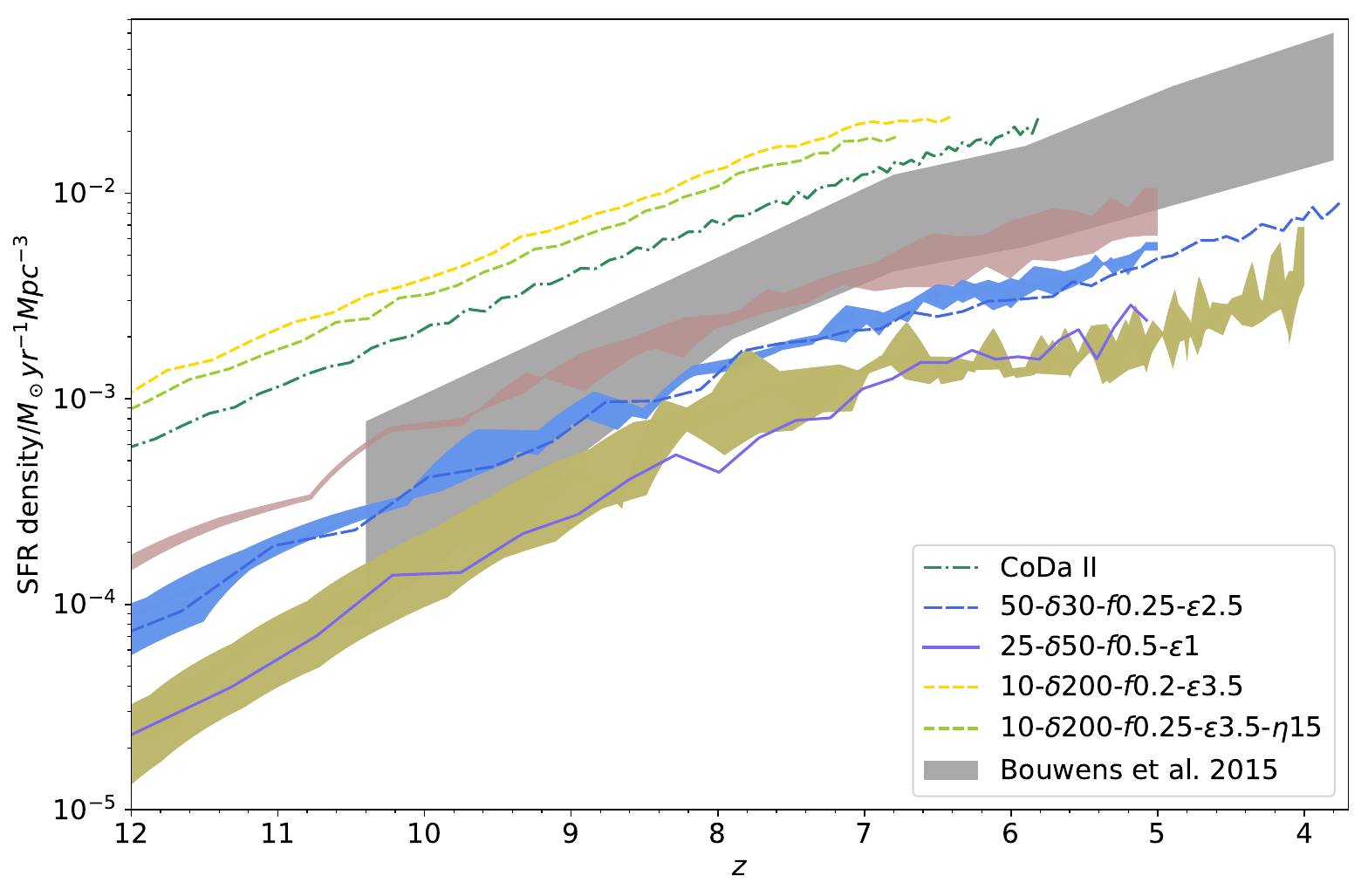}
  \caption{
    Evolution of the global star formation rate density. The dust-corrected and
    dust-uncorrected observations from \citet{Bouwens:2015}
    are indicated by the grey shaded      region.
    The yellow band includes the results from the four 100 Mpc boxes, 
    the rose-brown band indicates the range for early-reionization $25\,h^{-1}$Mpc box
    models, and the blue band includes the late-reionization ones.
%
%
%
%
	\label{fig:calib_plots_SFR}}
\end{figure}

In Figure~\ref{fig:calib_plots}, the top 2 panels show the evolution of the mean neutral and
ionised hydrogen fractions, respectively, along with the corresponding observational
constraints. The observational data
suggest that reionization is largely complete by $z\sim6$, at which point the mean neutral
fraction sharply drops to $\langle x_{HI}\rangle\lesssim 10^{-4}$ as the ionized patches overlap.
The universe becomes largely transparent, and the neutral fraction continues to decrease, but
more slowly. This sharp transition marks the end of reionisaton. Our simulations follow the
same general trend.

For clarity, we have grouped the reionisation histories of the four 25~Mpc
boxes that reionise earlier in the rose-brown band,
while the blue band represents the four 25 Mpc simulations that reionize later. 
The two high-resolution, small-volume simulations (dashed lines) were not run all the way to
completion of reionisation, but only to $z\sim6.3$ (\tenA) 
and $z\sim6.7$ (\tenB), but both follow the same general trend. 

The overlap epoch suggested by data is best matched by CoDa~II, \hunB, \hunC,  \fifty and \simH.
The additional simulations, which reionise slightly earlier and later than this allow us to
probe the effects of the timing of the EoR on the tSZ signal. We note that CoDa II, and to a
lesser extend the other early-reionization scenarios tend to over-ionize after overlap, while
the late-reionization models tend to agree with post-reionization data better. This does not
have significant effect on the tSZ signals, however.

\subsubsection{Global Photoionisation Rate}

In Fig.~\ref{fig:calib_plots}, bottom left panel we show the evolution of the mean global
photoionization rates.  
The evolution of the mean photoionisation rate roughly follows the same pattern for all
simulations, mirroring the reionisation histories. The early-reionization cases have
higher photoionisation rates, since their background UV flux densities are higher. More
ionising photons in the IGM allows for reionisation to proceed more rapidly in these runs.
Conversely, the simulations that reionise later have lower rates of photoionisation.
Around $z\sim6$, there is a noticeable upturn in the slopes. This corresponds to the end
of reionisation, where there is a drop in ionising photon absorption as the neutral hydrogen
in the boxes are ionised, increasing the background UV flux. While the slopes match the
pattern of the observational data, most of the simulations overshoot the observations
post-reionization. This discrepancy is responsible for those simulations having neutral
fractions lower than those observed. However, the blue band of 25 Mpc simulations which
best match with the neutral fraction observations also agree with the observed data for
the photoionisation rates. We also note that the photoionization rate evolution in the
higher-resolution simulations (CoDa II and the two 10~Mpc/h per side volumes) is
somewhat different, with initially low values, but a steeper rise around overlap.

\subsubsection{Optical Depth to Reionisation}

Another key observational constraint is the integrated Thomson scattering optical depth
to recombination, showing in Fig.~\ref{fig:calib_plots}, bottom left panel, along with the
corresponding result from Planck \citep{Planck-Collaboration:2018} 
and its 1-$\sigma$ errors (gray shared area). The Helium is not explicitly accounted for in our
simulations, thus its contribution is included as discussed in \citep{Iliev:2006}. assuming
that for $z>3$ He is singly ionized wherever the ionised fraction of Hydrogen is above 0.95, 
and for $z<3$ Helium is always double-ionized.

All cases are in rough agreement with the Planck results, at the low end, or just outside
the 1-$\sigma$ uncertainty. The late-reionization scenarios naturally have lower optical
depths, as there is less free electrons in the IGM along the line of sight between the end
of the EoR and the present day; and vice versa. The simulations whose optical depths best
agree with observations are \hunD\!\!, \simC\!, and \tenB \!\!.

\subsubsection{Star Formation Rate}

Finally, in Figure~\ref{fig:calib_plots_SFR}, we show the evolution of the global star
formation rate density. The grey shaded area shows the range between observed dust-corrected
and dust-uncorrected star formation rate (SFR) densities from \citep{Bouwens:2015}.
Here, the rose-brown band represents the set of higher star formation 25~Mpc volumes
(\simA,\simB,\simC),
while the blue band represents the lower star formation 25~Mpc volumes (\simD,\simE,\simF,\simG),
and the yellow band represents the 100~Mpc volumes. Similarly to the observational data, the
general trend of the SFR densities for all simulations is a continuous increase, with roughly
the same slopes. All simulations are in rough agreement with the SFR data, particularly most
25~Mpc volumes, the 50 Mpc one and CoDa II. The timing of reionization has little correlation
with the SFR.

The variation in the SFR of the simulations is mainly governed by the parameters $\delta_*$
and $\epsilon_*$. Since $\delta_*$ indicates the minimum gas over-density required to
trigger star formation, its value is tied to the spatial resolution of the box. Setting
$\delta_*$ higher results in fewer stars being formed since the (fixed-grid) simulation is
less able to resolve the high-density gas clouds. On the other hand, low $\epsilon_*$ values
mean that smaller fraction of the gas is converted into stars, resulting in lower SFR. Hence,
the balancing act between these two parameters is needed for obtaining a SFR corresponding to
the observational data. Unlike the previous plots, the SFR by itself does not directly indicate
the progression of the EoR, i.e. the relationship between them is not as straightforward. The
amount of ionising photons emitted into the IGM by the stellar population is dictated by the
value of $f_{esc,*}$. This parameter can be adjusted to ensure that the timing of the EoR
coincides with observations. Thus, it is possible for a simulation to have a low SFR but still
complete reionisation at a reasonable time. For example, \simH
has one of the lowest SFRs, due to its high $\delta$ and low $\epsilon_*$, but because its
assumed $f_{esc,*}$ is quite high, a greater proportion of photons are emitted into the IGM
per stellar particle, increasing the photoionisation rate, and allowing reionisation to end
at $z\sim6$. Another example of how the SFR history alone
does not govern the timing of the EoR is the fact that the simulations which have similar
SFR histories (grouped together in the three coloured bands) end reionisation at different
times. Within each coloured band, there are simulations which finish reionising as early
as $z\sim6.5$ and those which complete reionisation as late as $z\sim5.5$. Again, this is
a direct result of the different $\delta_*$, $\epsilon_*$ and $f_{esc,*}$ parameters used for them.
We see that increasing $\eta_{SN}$ by 5\% has only a small influence on the EoR, and a negligible
change in the SFR. \simD  \,
  completes reionisation slightly sooner ($z\sim5.8$) than does \simE   
 ($z\sim5.7$), and their SFR density histories are almost identical. Hence,
this small increase in supernova feedback only minimally influences the reionisation process.
We also note the high SFR of CoDa II. In addition to the different values used for
$\delta_*$, $\epsilon_*$, and $f_{esc,*}$, CoDa II has different simulation settings compared to the others, particularly the switching off of the temperature threshold for star formation. By turning off this setting, all cells, regardless of their temperature, are eligible for star
formation if they are above the density threshold. This means that the conditions for star formation are solely based on $\delta_*$ and $\epsilon_*$. This results in there being more regions where star formation is possible. While radiative suppression of star formation
still occurs in CoDa II, it is not as intense as it would be if the temperature threshold was turned on.


\section{Method}\label{sec:method}

\subsection{Electron Pressure Lightcones}\label{sec:electron_pressure_lightcones}

The input data for our calculations are the full-box, reduced-resolution cubes of the
volume-weighted gas pressure and mass-weighted ionisation fraction (CoDa~II) and the
full-resolution cubes of the gas pressure and ionisation fraction (the rest of our
simulations). These fields are then used to calculate the Compton y-parameter
(Eq.~\ref{eq:y-param}). We start by constructing lightcones of the electron
pressure, which for pure hydrogen gas is given by
\[
p_{\mathrm{e}} = \frac{x_{\mathrm{HII}}}{1 + x_{\mathrm{HII}}} p_{\mathrm{gas}}.
\]

We construct the lightcone as follows. Consider a photon travelling along a line of
sight, cell by cell, from initial redshift, $z_i$, to a final one, $z_f$. The age of
the universe, initial time and redshift can be readily calculated \citep[see e.g. Equation
  30 in][]{Hogg:1999}. We then convert the cell size from comoving to physical length by
using the corresponding scale factor $a$, and thus calculate the time it will take the
photon to cross the cell, $dt$. The age of the universe after this cell crossing is now
$t = t_1 + dt$. Assuming an Einstein-de Sitter universe, which is a good approximation
at high redshift, the photon is now at redshift
$z = \left(1 + z_i\right) \left({t}/{t_i}\right)^\frac{2}{3} - 1$.
We can now calculate the electron pressure of the cell at $z$ by interpolating the data
at the snapshots on either side of this redshift, say $z_1$ and $z_2$ corresponding to
snapshots 1 and 2, respectively.

The electron pressure is interpolated using a Sigmoid function
\begin{equation}
    g = \frac{1}{1 + \me{-\beta z_p}},
\end{equation}
with $\beta = 2$ and $z_p = -10 + 20 \left( \frac{z - z_2}{z_1 - z} \right)$, as follows
\begin{equation}
    p_{ei} = (1 - g) p_{e1} + g p_{e2},
\end{equation}
where $p_{e}$ is the interpolated electron pressure at $z$, and $p_{e1}$ and $p_{e2}$ are
the electron pressure of the cell at $z_1$ and $z_2$ (snapshots 1 and 2), respectively.
We then repeat the process, calculating the cell size, the time taken for the photon to
cross it, and so on at the updated redshift $z$. This allows us to obtain a lightcone of
the electron pressure: a 3D grid of the field where each slice in the direction of light
propagation reflects the state of the universe at that moment in time.

We then perform an integration along the light travel path in order to obtain the
Comptonisation parameter $y$. For each box light crossing, we numerically integrate
the pressure over all lines of sight along each axis using the composite Simpson's rule.
For the integral we use the previously calculated time intervals between each cell in the
lightcone ($dt$). In order to avoid artificial amplification of the $y$-parameter due to
structural repetition during the box replication after each light crossing, the volume
was randomly shifted and rotated during the integration stage. This procedure yields a
2D grid of the $y$-parameter.

We create lightcones for the same range of redshift for each simulation, $z \sim 6$ to $12$,
corresponding to the duration of the bulk of EoR. An exception to this are the 10 cMpc boxes,
which were not run all the way to redshift 6, and thus we constructed lightcones between
$z \sim 12$ and their last available snapshots, ($z \sim 6.38$ for \tenA and $z \sim 6.74$
for \tenB). The number of light crossings required to construct the light cones vary from
$\sim 10.8$ (100 cMpc boxes), to 17.2 light crossings (CoDa~II), and up to $\sim 88.4$ light
crossings for \tenB and $\sim 97.8$  for \tenA. 


We present the results in the form of maps of the $y$-parameter, probability density functions
of $y$ and angular power spectra of the signal. The maps are simply the images of the 2D
$y$-parameter grids obtained from the integration. We consider the full resolution results
as well as with smoothed maps with Gaussian beams with FWHM of 1.2 arcmin and 1.7 arcmin.
These beams correspond to the resolution of the South Pole Telescope (SPT)\footnote{\url{https://pole.uchicago.edu/}} \citep{Carlstrom:2011} at 150 GHz and 95 GHz, respectively.
We do not use the SPT 220 GHz beam as the tSZ effect disappears near this frequency.
We smooth the data by converting the angular resolutions of the beams to comoving Mpc at the
redshift of the lightcone at the end of each light crossing. We then calculated the number of
cells spanning this size and use this value as our FWHM for the Gaussian beam. Since angular
sizes vary negligibly at the redshifts considered here, we ignore this effect when constructing
the lightcones.

We also consider the contribution to the $y$-parameter for each redshift of the lightcone
in order to determine the period of dominant contribution. For these PDFs, we calculate the
distribution of the $y$-values for the start of the lightcone to each redshift, i.e. from
$z \sim 12$ to 11, $z \sim 12$ to 10, etc.

For comparison with previous results and data we compute the angular power spectra of the maps
and show them along with the \citet{Shaw:2010} power spectrum template scaled to fit the
cosmological parameters as done by \citet{George:2015}, and the SPT observations of the total
CMB power spectrum at the 95 GHz and 150 GHz bandpowers.

\subsection{Separate contribution lightcones}
\label{sec:separate_contributions}

We also investigate the separate contributions to the $y-$parameter and its fluctuations
from each constituent field -- gas density, temperature and ionisation fraction. We do
this as follows, using the volume-weighted gas density and mass-weighted ionisation
fraction.

The fluctuations in the temperature field were removed by using a fixed global value
for the temperature, namely $T_{\mathrm{e}} = 30,000$ K. We then constructed an electron
density lightcone via the same steps as for the electron pressure lightcones in
\S~\ref{sec:electron_pressure_lightcones} for the redshift range $z \sim 6$ to 12,
and using the expression
\begin{equation}
    \rho_{\mathrm{e}} = \frac{x_{\mathrm{HII}}}{1 + x_{\mathrm{HII}}} \rho_{\mathrm{gas}}
\end{equation}
for the electron density. After the lightcone was constructed, we numerically
integrated the electron density over all lines of sight along each axis and along
the direction of light propagation, as before. The lightcone integral then becomes
\begin{equation}\label{eq:y_rho}
    y = 
    \frac{\sigma_\mathrm{T} k_{\mathrm{B}}}{m_\mathrm{e} m_{\mathrm{p}} c}
    T_{\mathrm{e}}
    \int \frac{x_{\mathrm{HII}}}{1 + x_{\mathrm{HII}}} 
    \rho_{\mathrm{gas}} 
    \ \mathrm{d} t,
\end{equation}
where $m_\mathrm{p}$ is the proton mass and comes from converting the electron mass
density to number density
$\left( n_\mathrm{e} = n_\mathrm{HII} = \frac{\rho_\mathrm{HII}}{m_\mathrm{p}} \right)$.
We also randomly shifted and rotated the box to avoid artificially boosting the
signal. Again, this yields a 2D grid of $y$-parameter values for our volume.

Next, we removed the patchiness of the EoR in addition to the temperature fluctuations,
by using a globally averaged value of the ionisation fraction for each light crossing.
We made a lightcone of gas density using only the volume-weighted gas density files
for CoDa II for the interpolation process, for the redshift range $z \sim 6$ to 12.
In this case, Equation~\ref{eq:y_rho} becomes
\begin{equation}
    y = \frac{\sigma_\mathrm{T} k_{\mathrm{B}}}{m_\mathrm{e} m_{\mathrm{p}} c}
    T_{\mathrm{e}} \left< x_\mathrm{HII} \right>
    \int \rho_{\mathrm{gas}} 
    \ \mathrm{d} t,
\end{equation}
where $\left< x_\mathrm{HII} \right>$ is the globally averaged ionisation fraction.
We used the value of $\langle x_{\mathrm{HII}} \rangle$ 
corresponding to the redshift of the box at the end of each light crossing.
Once again, we randomly shifted and rotated the box during integration, obtaining
a 2D map of the $y$-parameter.

\subsubsection{Phase Diagrams}
In order to further probe the underlying quantities contributing to the $y$-parameter,
we also construct and compare the phase diagrams of the gas temperature and density.
Since the electron pressure is dependent on these quantities, this allows for better
understanding of the trends and dependence on different paraneters. We present these
phase diagrams and an interpretation of them with respect to the $y$-parameter values
in Section~\ref{sec:2D_results}.

\subsection{Additional Corrections and Tests}
\subsubsection{y-parameter for Uniform IGM}
\label{sec:analytic}

If we assume instantaneous reionization, a constant post-reionization IGM
temperature and uniform density at the mean value for the universe, then the
y-parameter mean value can be calculated analytically. While clearly
unrealistic, this yields an useful reference value for our more detailed
results later.


We must first rewrite the integral with respect to redshift, since we are concerned with
the redshift evolution of the IGM density, The Friedmann equation:
\begin{equation}
  H^2 
  = H_0^2 
    \left[
    \Omega_{\mathrm{r},0} \left(1+z\right)^4 + \Omega_{\mathrm{m},0} \left(1+z\right)^3 + 
    \Omega_{\mathrm{k},0} \left(1+z\right)^2 + \Omega_{\Lambda,0}
    \right].
\end{equation}
for a flat universe, $\Omega_{\mathrm{k},0} = 0$, and at high redshift, $\Omega_{\mathrm{r},0} \sim 0$ and
$\Omega_{\mathrm{m},0} \gg \Omega_{\Lambda,0}$ reduces to
\begin{equation}\label{eq:Friedmann_analytic}
    \left( \frac{\dot{a}}{a} \right)^2 = H_0^2
    \left[
    \Omega_{\mathrm{m},0} \left(1+z\right)^3
    \right].
\end{equation}
Then, Equation~\ref{eq:y-param} yields
\begin{equation}
    y = - \frac{\sigma_\mathrm{T}}{m_\mathrm{e}c^2} \frac{1}{H_0 \Omega_{\mathrm{m},0}^\frac{1}{2}} \int p_\mathrm{e} \left(1+z\right)^{-\frac{5}{2}} c \mathrm{d}z.
\end{equation}
Assuming the IGM is an ideal gas, $p_\mathrm{e} = k_\mathrm{B} n_\mathrm{e} T_\mathrm{e}$,
where $n_\mathrm{e}$ and $T_\mathrm{e}$ are the electron number density and temperature, respectively.
Also, for pure-hydrogen gas we have
\begin{equation}
  \label{el_dens}
    n_\mathrm{e} = n_\mathrm{HII} = \frac{\rho_\mathrm{HII}}{m_\mathrm{HII}}
\end{equation}
and
\begin{equation}
    \rho_\mathrm{HII} = \rho_\mathrm{gas} \frac{x_\mathrm{HII}}{1+x_\mathrm{HII}}.
\end{equation}
Since $m_\mathrm{HII}$ is simply the proton mass, we shall use the notation $m_\mathrm{p}$ henceforth.
For instantaneous reionisation, $x_\mathrm{HII}$ can only be either 0 or 1 if neutral or ionised, respectively.
Hence, $\frac{x_\mathrm{HII}}{1+x_\mathrm{HII}} = \frac{1}{2}$ for a reionised IGM.
Thus the electron number density for uniform IGM is 
\begin{equation}\label{eq:ne}
  n_\mathrm{e} = \frac{1}{2} \frac{\rho_\mathrm{gas}}{m_\mathrm{p}}
  = \frac{1}{2} \frac{\rho_0 \Omega_{\mathrm{b},0} \left( 1+z\right)^3}{m_\mathrm{p}}, 
\end{equation}
where $\rho_0 = 10^{-30}$ g cm$^{-3}$ is the mean density of the universe at $z = 0$.
Converting the integrand from electron pressure to density for an ideal gas, we get
\begin{equation}
    y = - \frac{1}{2}
    \frac{\sigma_\mathrm{T}k_\mathrm{B}}{m_\mathrm{e}m_\mathrm{p}c} 
    \frac{\rho_0 \Omega_{\mathrm{b},0}}{H_0 \Omega_{\mathrm{m},0}^\frac{1}{2}}
    T_\mathrm{e}
    \int
    \left(1+z\right)^\frac{1}{2}
    \mathrm{d}z.
\end{equation}

Evaluating this integral for the redshift range of the lightcones ($z = 12$ to $6$) and the
cosmological parameters for CoDa II, and assuming an average IGM temperature of $30,000$ K,
we obtain a Comptonisation parameter $y \approx 4.22 \times 10^{-8}$. 
For the cosmology
of the other simulations, we find a very similar mean value of $y \approx 4.18 \times 10^{-8}$.

\subsection{Helium Reionisaton}
\label{sec:He-reion}

So far, we have assumed that the baryonic matter is comprised of hydrogen only, ignoring
helium and metals. While the presence of metals is insignificant to this study, the presence
of helium will contribute to the tSZ signal. Neutral helium (He~) requires photons with
energy of at least 24.6~eV to be singly ionised to He~II, with He~II recombining at roughly
the same rate as H~II. However, He~II requires photons of at least 54.4~eV energy to fully
ionise He III, with a recombination rate over 5 times that of hydrogen. Thus, when
accounting for the presence of helium, the sources that ionise hydrogen are often considered
to singly ionise helium. On the other hand, the second ionisation of helium occurs at later
redshifts, with the reionisation of helium ending at $z\sim3$ \citep[e.g.][]{Barkana2001}.
Therefore, during the redshift range considered in this study, we assume that each helium
atom contributes one electron to the IGM whenever hydrogen is ionized. Equation~\ref{el_dens}
thus becomes
\begin{equation}
n_e = n_{\rm HII} + n_{\rm HeII} ,
\end{equation}
where $n_{\rm HeII}$ is the number density of singly-ionised helium. When accounting for doubly-
ionised helium, the term $2n_{\rm HeIII}$ is added.

Assuming the primordial abundances of hydrogen and helium by mass ($X = 0.76$ and
$Y = 0.24$, respectively), and ignoring any isotopes, we get
\begin{equation}
  ne = x_{\rm HII}X\frac{\rho_{gas}}{m_H} + x_{\rm HeII}Y\frac{\rho_{gas}}{4m_H}
=\left(x_{\rm HII}X+x_{\rm HeII}\frac{Y}4\right)\frac{\rho_{gas}}{m_H}
\end{equation}
where $x_{\rm HeII} = n_{\rm HeII}/n_{\rm He}$ is the ionisation fraction of singly ionised helium. For
instantaneous reionisation, we have $x_{\rm HII}$ and $x_{\rm HeII}$ jump from 0 to 1 at the
transition, and we have
\begin{equation}
n_e = 0.82\frac{\rho_{gas}}{m_H}
\end{equation}

This results in an 18\% decrease in the tSZ signal, compared to the hydrogen-only numbers.
The estimated mean Comptonisation parameters for CoDa II and the auxiliary simulations
therefore become $y\approx 3.46\times10^{-8}$ and $y \approx 3.43 \times 10^{-8}$,
respectively. In addition, RAMSES-CUDATON only tracks the ionisation and cooling processes 
for atomic hydrogen. In doing so, it assumes that hydrogen makes up 76\% of the baryonic
matter, thus preserving its primordial abundance. Nevertheless, the temperature field
generated by the simulation is given in units of $K\mu^{-1}$, where $\mu$ is the mean
molecular weight of the gas. Since $\mu$ depends on the ionisation fractions,
\begin{equation}
\frac1{\mu}= (1 + x_{\rm HII})X + (1 + x_{\rm HeII} + 2x_{\rm HeIII})\frac{Y}4,
\end{equation}
  the temperature field (and, hence, the pressure field) can be updated, in post-processing,
to roughly account for helium reionisation, by interpolating $\mu$ values for the required
redshift range.

\subsection{Quadratic (second-order) Doppler contribution}
We estimate the contribution to the $y$-parameter due to the second-order Doppler distortions using a similar methodology to the one used for calculating the mean Compton-y from tSZ effect. We start by isolating the $\mathcal{O}(v^2)$ term in Equation~\ref{eq:yso} and recast the integral to be dependent on $z$
\begin{equation}
    y = \frac{\sigma_{\rm T}}{3 c H_0} \int^{z_1}_{z_0} \frac{n_{\rm e} \langle v^2 \rangle}{(1+z)E(z)}  {\rm d}z
    \label{eq:ysoz}
\end{equation}
where $n_{\rm e}$ and $\langle v^2 \rangle$ are extracted from the simulations, as follows.
For each cell, we first compute 
\begin{equation}
    n_{\rm e} = \frac{\rho_{\rm HII}}{m_{\rm HII}}=\frac{X\rho x_{\rm HII}}{m_{\rm p}} 
\end{equation}
and 
\begin{equation}
    v^2 = v^2_x + v^2_y + v^2_z,
\end{equation}
where we have dropped the angled brackets for clarity, before taking the average over a snapshot to obtain $\langle n_{\rm e}v^2 \rangle$ for a given $z$. With these values of $\langle n_{\rm e}v^2 \rangle$ as a function of $z$ in hand, we numerically integrate Equation~\ref{eq:ysoz} between $z=6$ and $z=12$, linearly interpolating between the available snapshots. We confirm that the result is converged with respect to the interpolation sampling.

The power of the velocity field peaks at fairly large scales $\sim100~$Mpc \citep[for more detailed discussion see e.g.][]{Iliev:2007}, thus a significant fraction of that power is missing in our relatively small simulation volumes. However, at large scales the velocity field is linear and thus the effect of this missing power can be readily estimated and included in our calculations, as follows.   
Starting from the power spectrum of density fluctuations, defined as
\begin{equation}
    \langle \delta(\mathbf{k}, z) \delta(\mathbf{k'}, z) \rangle = (2 \pi)^3 \delta_{\rm D}(\mathbf{k} + \mathbf{k'})P(k, z),
\end{equation}
where $\delta(\mathbf{k}, z)$ is the Fourier transform of the real-space density fluctuations and $\delta_{\rm D}$ is the Dirac delta function, we use the continuity relation
\begin{equation}
    \mathbf{v}(\mathbf{k}, z) \approx \frac{i H(z) f[\Omega_{\rm m}(z)]\mathbf{k}}{(1 + z)k^2} \delta (\mathbf{k}, z),
\end{equation}
where $f[\Omega_{\rm m}(z)]\approx \Omega_{\rm m}(z)^{0.6}$ \citep{lahav1991}, to write the total mean-square fluctuations for the velocity in linear theory as
\begin{equation}
    \langle v^2\rangle_{\rm tot} = \frac{H(z)^2 f[\Omega_{\rm m}(z)]^2}{2\pi^2(1+z)^2} \int^\infty_0 P(k)  {\rm d} k.
\end{equation}
Finally, we estimate the mean-square velocity missing from a simulation box through
\begin{equation}
\langle v^2 \rangle_{\rm missing} = \langle v^2 \rangle_{\rm tot} -  \langle v^2 \rangle_{\rm box},
\end{equation}
where $\langle v^2 \rangle_{\rm box}$ is the mean-square velocity calculated directly from the simulation box. This missing power can then be added as a correction to the simulation results from sub-box scales.

\section{Results}\label{sec:results}

\subsection{Electron Pressure Lightcones}\label{sec:electron_pressure_lightcones_results}

\begin{table}
    \caption{The mean Compton $y$-parameter for each simulation.}
    \centering
    \begin{tabular}{|l|c|}
        \hline
        \textbf{Simulation} & $\mathbf{\left<y\right> \times 10^8}$ \\
        \hline
        CoDa II & 3.67 \\
        \hline
        \hunA & 1.22 \\
        \hunB & 1.96 \\
        \hunC & 1.91 \\
        \hunD & 2.70 \\
        \hline
        \fifty & 2.47 \\
        \hline
        \tenA & 2.62 \\
        \tenB & 2.46 \\
        \hline
        \simA & 2.86 \\
        \simB & 2.18 \\
        \simC & 3.63 \\
        \simD & 1.97 \\
        \simE & 2.03 \\
        \simF & 2.96 \\
        \simG & 3.09 \\
        \simH & 1.82 \\
        \hline
    \end{tabular}
    \label{tab:mean_y}
\end{table}

The main results of our study is the tSZ effect calculated from the electron pressure lightcones
calculated as discussed in Section~\ref{sec:electron_pressure_lightcones}. The mean $y$-parameter
values, $\langle y\rangle$, found from these lightcones for the sixteen simulations are listed in
Table~\ref{tab:mean_y}. All values of $\langle y\rangle$ we find are of order a few$\times10^{-8}$,
well below the COBE-FIRAS limit of $\langle y\rangle < 1.5 \times 10^{-5}$ \citep{Fixsen:1996},
and two orders of magnitude lower than the total mean $y$-parameter estimated by
\citet{Refregier:2000, Zhang:2004a, Hill:2015}. This is expected, since galaxy clusters provide
the dominant contribution to this quantity, at $\langle y\rangle_\mathrm{ICM}=1.58 \times 10^{-6}$
\citep{Hill:2015}. Compared with the reionisation contribution estimated by \citet{Hill:2015},
namely $\langle y\rangle_\mathrm{EoR} = 9.8 \times 10^{-8}$), our values are slightly lower, but of
the same order of magnitude. This discrepancy is most likely due to differences between our
simulations and their reionisation model. Nevertheless, these values for the mean Comptonisation
parameter for the EoR are comparable, albeit approximate models might sometimes overestimate the
mean EoR signal. As we will see below, the value of the y-parameter closely correlates with the
end of reionization redshift, with early overlap yielding higher y-values.

\subsubsection{The tSZ Signal and the Timing of the EoR}

\begin{figure}
	\includegraphics[width=1\linewidth]{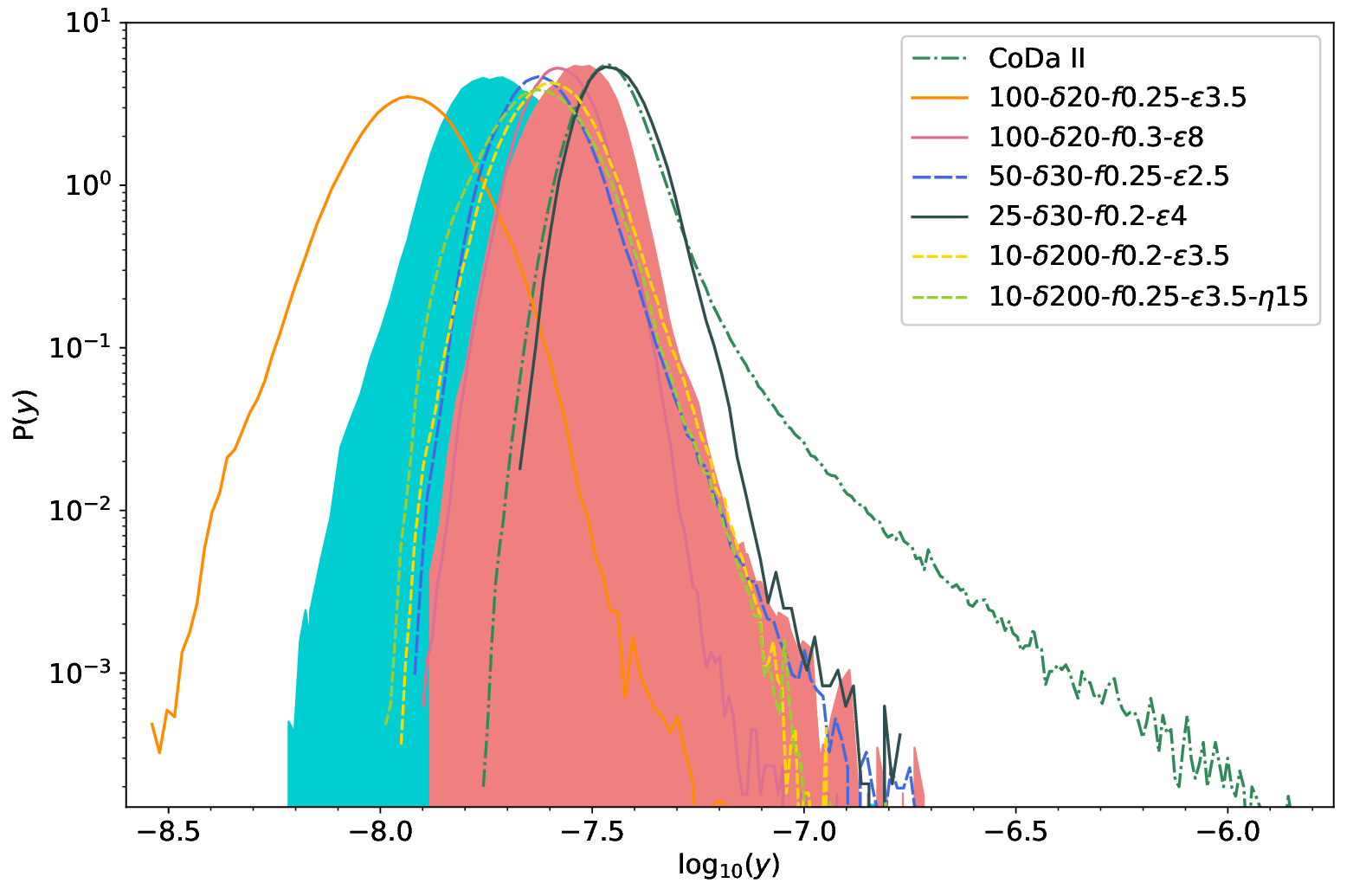}
	\caption{PDF distributions of the comptonization $y$-parameter for our simulations,
          some of which are shown individually (as labelled) and for clarity the rest are
          grouped in two bands - turquoise one including \hunB, \hunC, \simA, \simD, \simE
          \, and \simH, and coral one for \simB, \simF, and \simG.}
	\label{fig:pdf}
\end{figure}

For further insight and understanding of these values, we consider the full probability
density functions (see Fig.~\ref{fig:pdf}). For clarity, we only show a few of the
simulations individually (as labelled), and we group the rest into two bands - first
includes \hunB, \hunC, \simA, \simD, \simE\, and \simH \, (turquoise) and second - \simB,
\simF, and \simG \, (coral).
Simulations in the former group have lower y-parameter values ($y\lesssim2\times10^{-8}$)\,
and generally late end of reionisation. The only simulation with an even smaller tSZ signal
is \hunA,
which is also the case with the latest end of EoR.
The second band (coral) groups cases with higher y-values and early-finishing reionization. 

At the high-end of y-values are CoDa II and \simC, which is the simulation with the
earliest overlap and highest y-parameter values among the auxiliary simulations.
Their PDFs overlap, save for the long high-y tail in the case of CoDa II, which is
a result of its higher resolution, and different supernova feedback model. This
allows resolving small, high-temperature regions around supernova explosions, resulting
in small areas of very high pressure, and therefore bright in tSZ. However, we note that
despite their similar PDFs, the reionisation histories of those two simulations differ
significantly, with \simC \,\,
reionising considerably earlier ($z_{reion}\sim 6.6$) than CoDa II ($z_{reion}\sim6.1$).


We also note that earlier and faster reonization correlates with narrower y-parameter
distributions, since such scenarios mean most of the gas being ionized at similar redshifts,
and the gas has more time to equilibrate after overlap, resulting in less fluctuations in the
ionised hydrogen fraction and more narrow range in y-values. These results indicate a link
between the timing of the EoR and the strength of the tSZ effect stemming from that cosmic
period. When reionisation ends earlier, there is a longer period of fully-ionized gas, thus
higher y-parameter overall.

\begin{figure}
    \centering
    \includegraphics[width=1\linewidth]{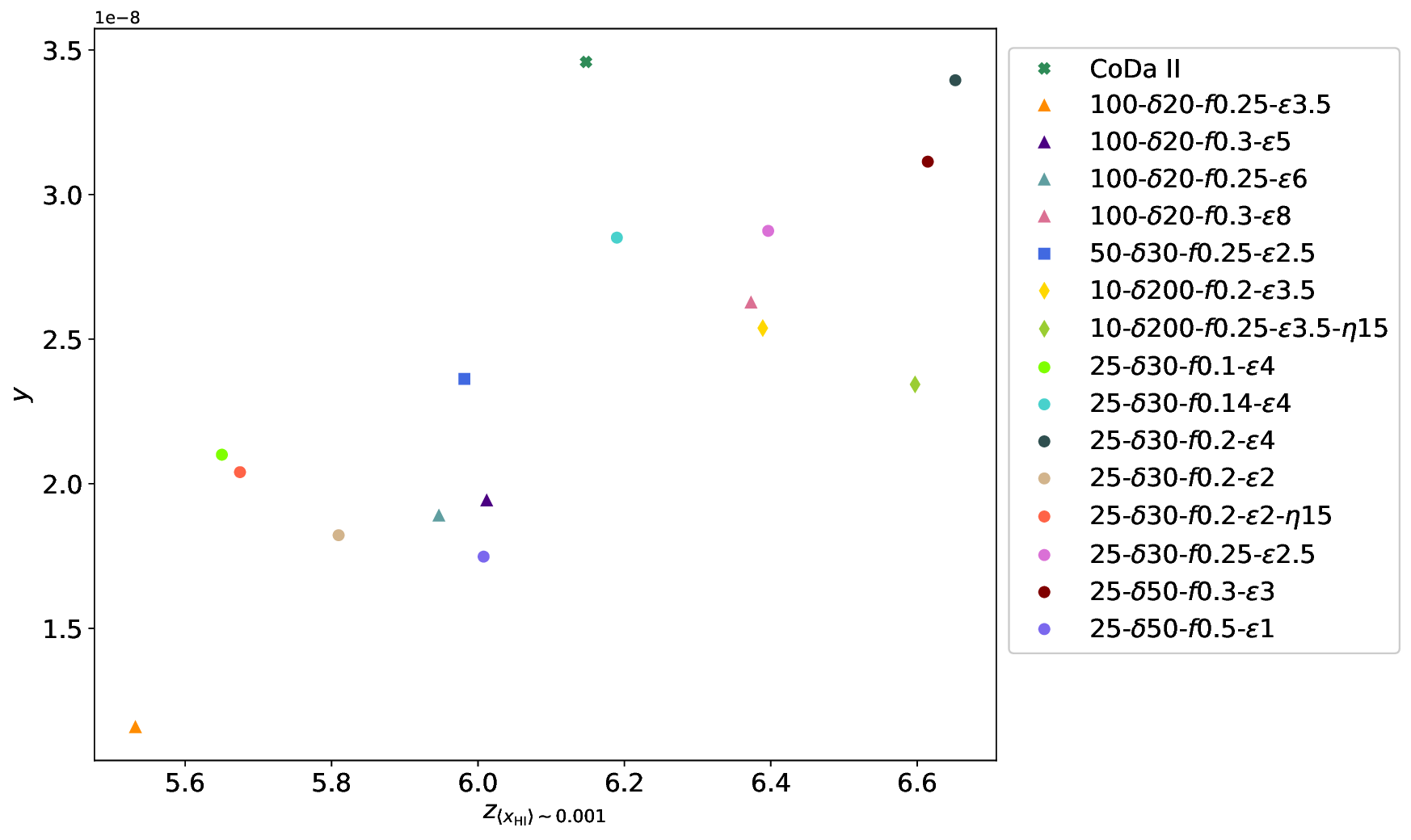}
    \caption{The mean y-parameter for each simulation (as indicated in the legend) vs.
      $z_{\langle x_{HI}\rangle\sim0.001}$, the redshift at which $x_{HI}\sim0.001$.}
    \label{fig:pdf_y_z}
\end{figure}

In order to better visualise the relationship between the tSZ effect and the timing
of the EoR, in Fig.~\ref{fig:pdf_y_z} we show the mean y-parameter vs. the redshift at which the
average neutral hydrogen fraction reaches $10^{-3}$, referred to as $z_{\langle x_{HI}\rangle\sim0.001}$.
We see that there is indeed a clear positive correlation between $\langle y\rangle$ and
$z_{\langle x_{HI}\rangle\sim0.001}$, albeit with significant scatter. Curiously, the most significant
outliers, namely CoDa II and \tenB, are among the best-resolved simulations in our set.
However, we do not believe this is a resolution effect, since these two outliers are at opposite
extremes, while our other high-resolution simulation, \tenA, is exactly on the trend, as are the
rest of our simulations, despite their varying resolutions. For CoDa II, this is most likely due
to the different simulation settings used in that run, while 
the lightcone for \tenB \,\,  has the shortest redshift range, resulting in lower mean y-parameter.

In addition to their different EoR timings, these boxes have varying SFR histories, which
also influences the observed scatter. For two simulations with similar y-parameters but
different EoR timings, the SFR of the simulation with a later EoR is higher than that of
the one whose EoR ends earlier. This suggests that, although a simulation may reionise later,
if the corresponding SFR is relatively high, the tSZ effect is boosted. A higher SFR leads
to more stars which then go supernova and raise the local gas temperature and pressure. The
electron temperature therefore increases, boosting the inverse Compton scattering.

\begin{figure*}
    \centering
    \includegraphics[width=1\linewidth]{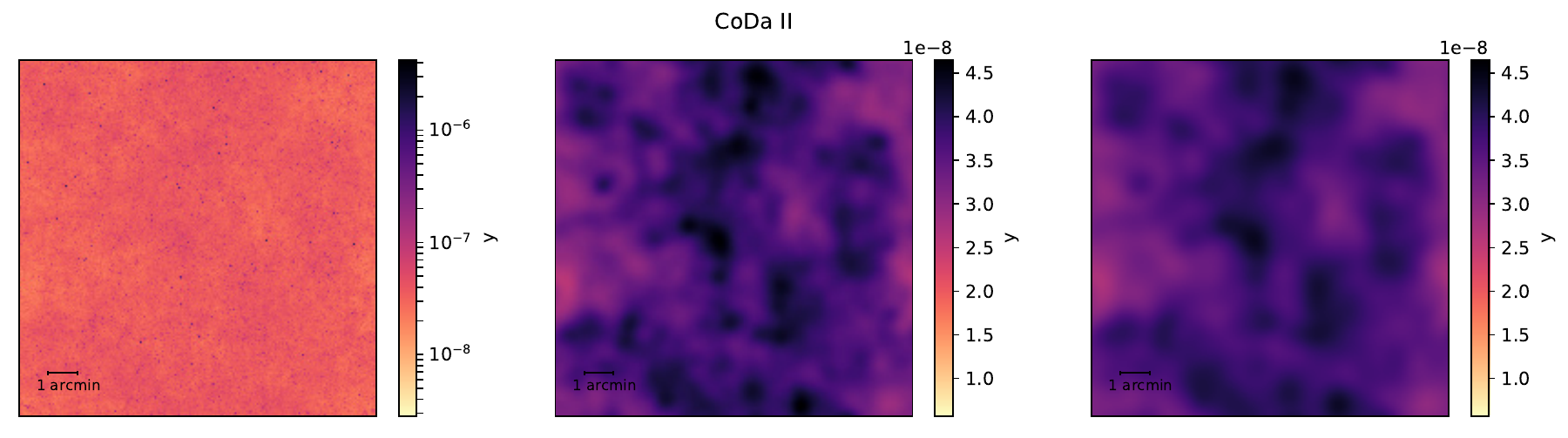}
    \includegraphics[width=1\linewidth]{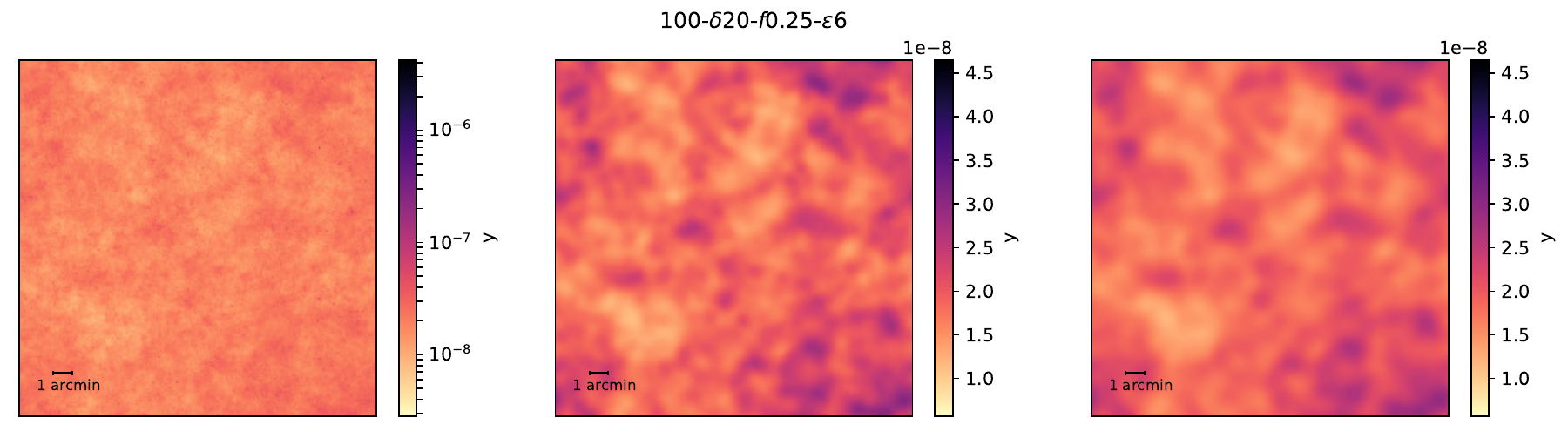}

    \centering
    \includegraphics[width=1\linewidth]{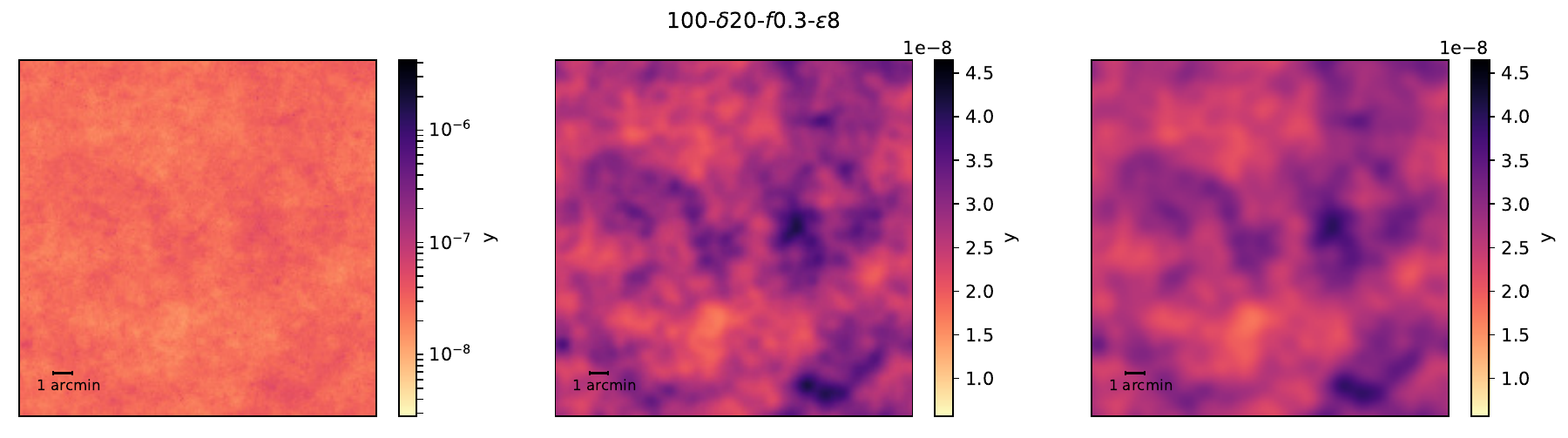}
    \caption{Maps of the Comptonization $y$-parameter for CoDa II, 100 \C, and 100 \D \,\,  simulations. 
    \textit{Left:} full resolution maps sharing the same colour bar scale for $y$. \textit{Middle:} maps smoothed with a Gaussian beam of FWHM 1.2 arcmin FWHM, corresponding to the resolution of the 150 GHz channel of the SPT. \textit{Right:} maps smoothed with a Gaussian beam of FWHM 1.7 arcmin, corresponding to the 95 GHz channel of the SPT. Smoothed maps share the same colour bar scale.}
    \label{fig:maps1}
\end{figure*}

\begin{figure*}
    \centering
        \includegraphics[width=1\linewidth]{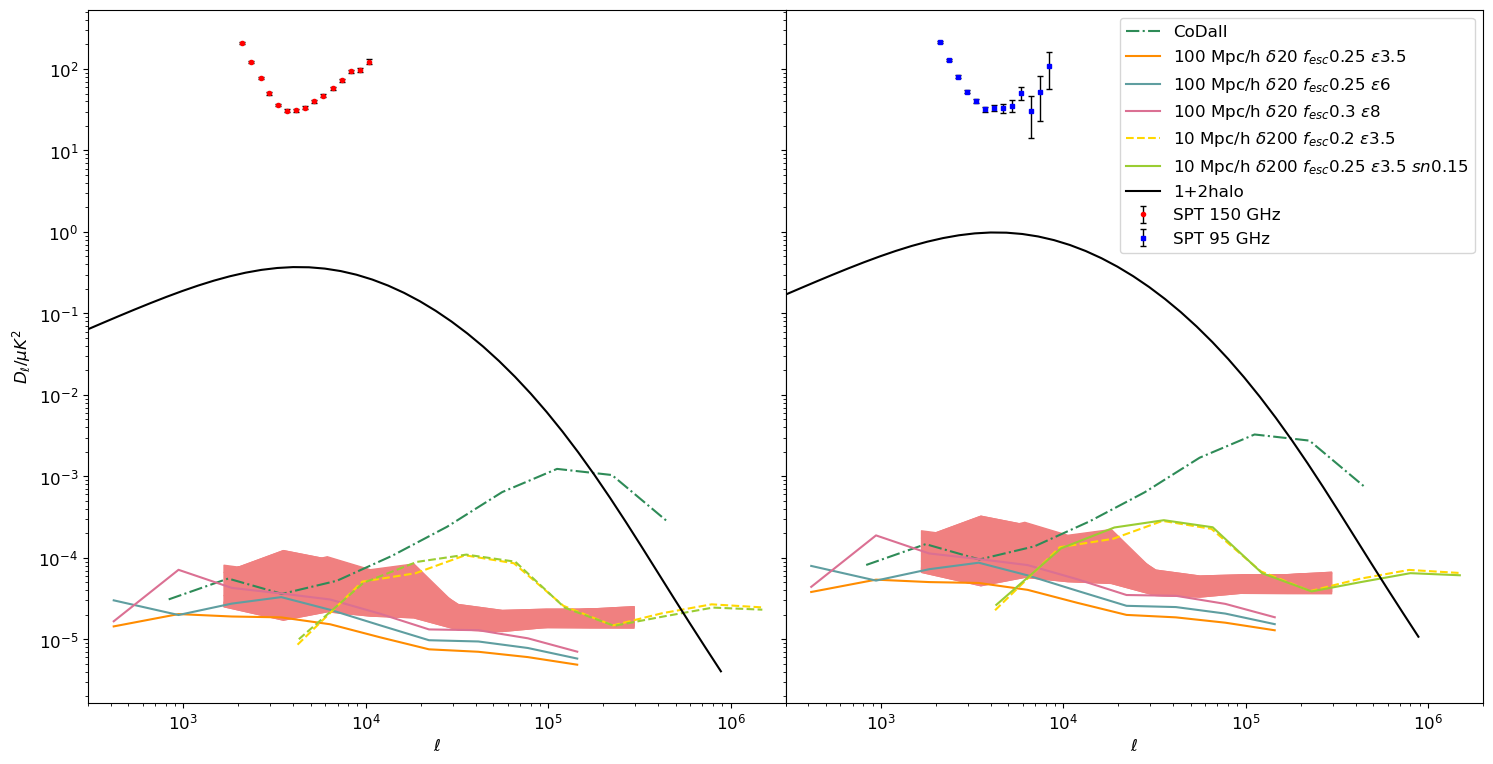}
    \caption{The angular temperature power spectra of the tSZ signals yielded by oour simulations. Power spectra calculated for the two SPT 
    band-powers in which the tSZ effect is visible: 150 GHz (\textit{left}) and 95 GHz (\textit{right}). The black solid line is the 1-halo 
    and 2-halo combined contribution of the theoretical \(y\)-power spectrum by \citet{Bolliet:2018}, while the points with error bars are
    the SPT data for the total CMB power spectrum at the respective frequencies \citep{George:2015}.}
    \label{fig:aps}
\end{figure*}

\subsubsection{Impact on Galaxy Cluster Measurements}


Another objective of our study is to probe the effect of the EoR tSZ signal on galaxy
cluster measurements. Although the mean tSZ signal from the EoR is subdominant to that 
of clusters, this might not be the true at all scales. Furthermore, even if sub-dominant,
it is useful to  quantify how much error the former might contribute toward the latter.
We start by showing sample Compton y-parameter maps from several of our simulations in 
Fig.~\ref{fig:maps1} 
and the angular power spectra for all simulations in Fig.~\ref{fig:aps}. In Fig.~\ref{fig:maps1} 
we present the full-resolution map (left panels) as well as maps smoothed with Gaussian beams of 
FWHM of 1.2 arcmin (middle panel) and 1.7 arcmin (right panel), corresponding to those of the 150 GHz
and 95 GHz SPT channels, respectively. The fluctuations in the
full-resolution maps are strongest in the sub-arcminute scale, with the largest y-parameters
reaching $y\sim10^{-6}$. When the maps are smoothed, these fluctuations are largest at 
$\sim1~$arcmin, with values of order $y\sim\,\textrm{few}\times10^{-8}$ . On the other hand, 
the y-parameter of galaxy clusters is typically $y<10^{-5}$, and their angular size ranges 
from from tens of arcsec to tens of arcmin. From the power spectra, we see that, at cluster 
scales ($l<10^4$), the EoR accounts for < 1\% of the cluster signal, but could be significantly
larger fraction at smaller scales. These observations are more quantitatively confirmed by the 
angular power spectra from our full-resolution maps  

The angular power spectra of the full-resolution maps shown in Fig.~\ref{fig:aps} along with 
the \citet{Bolliet:2018} 
tSZ template (1-halo+2-halo terms;  solid black line) and the total CMB power spectrum measured 
by the SPT \citep{George:2015}. Our predicted sky power spectra are flatter in shape than the 
\citet{Bolliet:2018} 
template. On smaller scales ($l > 10^4$), the tSZ signals from CoDa II and the 10 Mpc boxes
become stronger, possibly surpassing the post-EoR signal. The other simulations have lower 
resolutions and do not allow for the signal at these small scales to be calculated.  
For the current instruments like SPT with an angular resolution of $\sim1$~arcmin, the areas
of highest electron pressure in the EoR produce a tSZ signal which is about three orders
of magnitude smaller than that of clusters. However, a more precise instrument, with
arcsec resolution and higher sensitivity, would be able to detect the small regions where
$y\sim10^{-6}$ in our full-resolution maps, and extend the power spectra to $l\sim10^5$,
where the EoR contribution becomes stronger, and possibly dominant. Although the EoR is a 
sub-dominant contributor to the total tSZ effect on larger scales, it nevertheless should 
be considered for precision cosmology.

\begin{figure}
    \centering
    \includegraphics[width=1\linewidth]{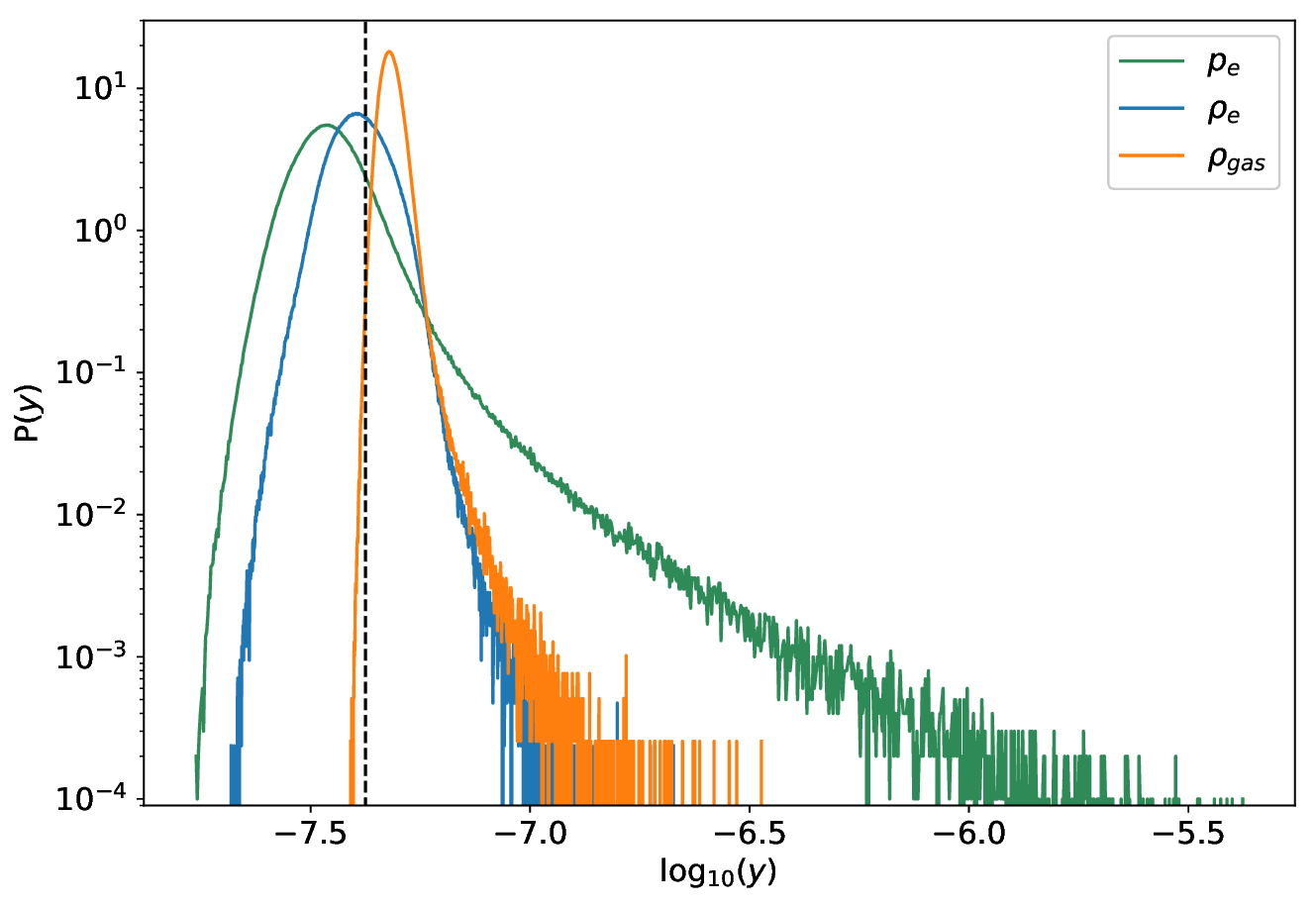}
    \caption{Distributions of the $y$-parameter for full CoDa II results (green), uniform
      global electron temperature (blue), both uniform global electron temperature and
      uniform ionized fraction (orange). Finally, the vertical black dashed line shows the
      analytical result for instantaneous reionization (\S.~\ref{sec:analytic},
      $\log_{10}y=-7.38$, $y=4.22\times10^{-8}$)}
    \label{fig:pdf_rho}
\end{figure}

\begin{figure*}
    \centering
    \includegraphics[width=0.49\textwidth]{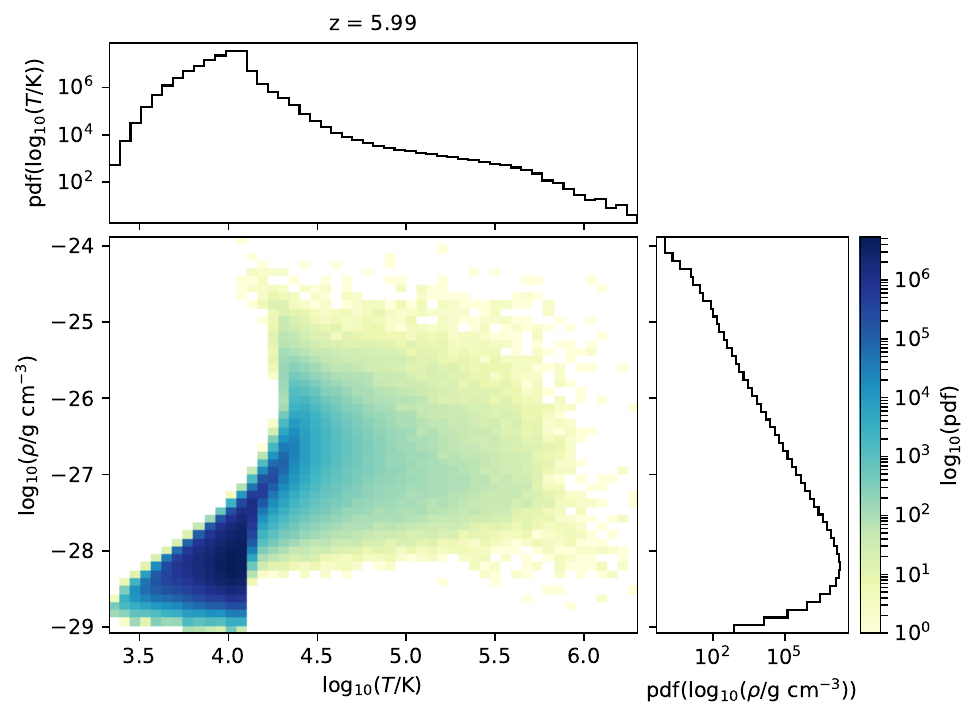}
        \label{fig:2D_CDII} 
        \includegraphics[width=0.49\textwidth]{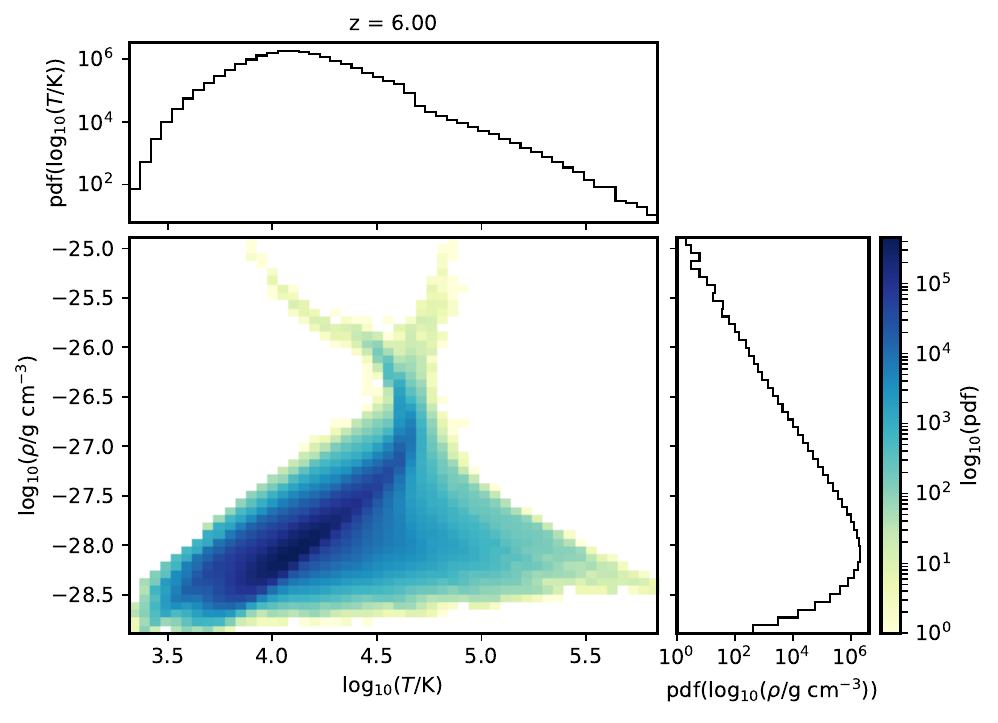}
        \label{fig:2D_hunC} 
        \includegraphics[width=0.49\textwidth]{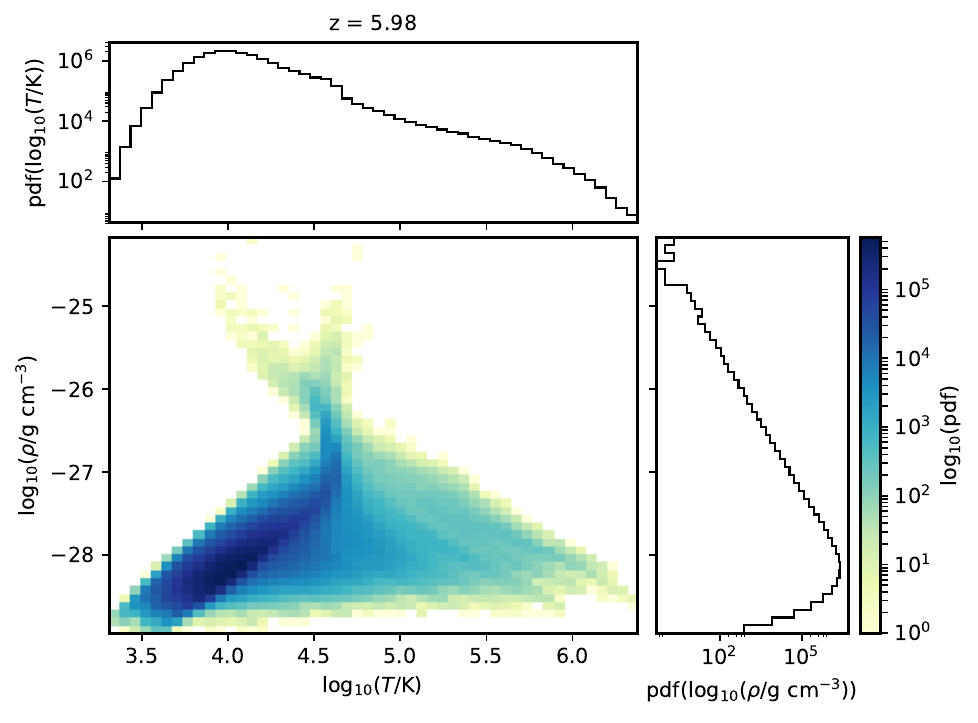}
        \label{fig:2D_fifty} 
        \includegraphics[width=0.49\textwidth]{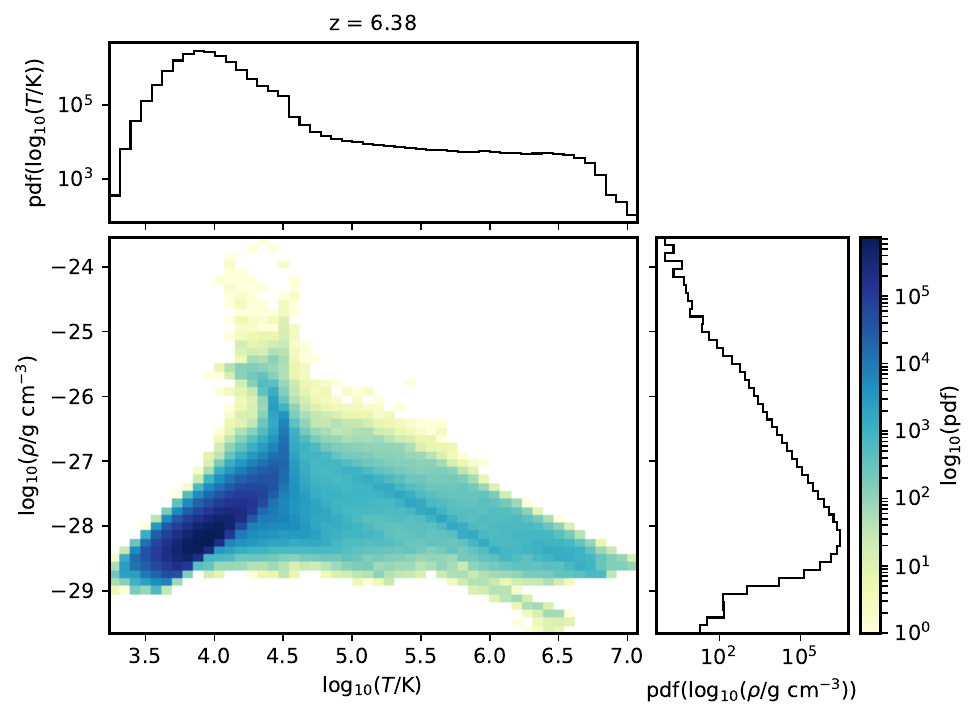}
        \label{fig:2D_tenA} 
        \caption{Phase diagrams showing the gas density vs. temperature at the end of the
          reionization ($z \sim 6$), for selected subset of our simulations: (top left)
          CoDa~II, (top right) \hunC, (bottom left) \fifty, and (bottom right) \tenA.}
    \label{fig:phase_diagrams1}
\end{figure*}

\subsection{Density Lightcones}
\label{sec:electron_density_lightcones_results}

The tSZ signal from the reionization epoch is formed by the complex interactions of a number
of factors and quantities, making it difficult to fully understand. In order to disentangle
and evaluate these different contributions to the signal, we next consider them separately,
as discussed in \S~\ref{sec:separate_contributions}. Specifically, we compare the full tSZ
results to series of simplified cases where we: 1) fix the electron temperature to
$T_e=30,000~K$ throughout the volume; 2) fix the electron temperature as well as the ionized
fraction to its globally-averaged value at that redshift; and 3) the analytic case in
Sec.~\ref{sec:analytic}, of instantaneous reionization and constant values of all fields.
All calculations are done based on the CoDa~II data and results are shown in Fig.~\ref{fig:pdf_rho}.

The peak of each distribution lies near to the analytic result, with the two patchy
reionization cases peaking slightly below it, and the uniform reionization case slightly
above it. Furthermore, the patchy reionization PDFs are wider, with significant fraction
of the lines of sight having low y-values, below $y=4.22\times10^{-8}$. In contrast, the
uniform reionization case yields much narrower distribution, with almost no lines of sight
below the analytical result. This is due to the presence of cold, neutral cells in the
patchy reionization cases, contributing little to the integrated Compton $y$ parameter.

Conversely, the high-y tail we observed in the full CoDa-II results is primarily due to
the highly-heated gas from SNe explosions and structure formation shocks, resulting in
local temperatures up to millions of K. When the electron temperature fluctuations are
removed (blue) this high-y tail completely disappears, leaving an almost Gaussian
distribution. When also the reionization patchiness is removed a small tail re-appears
at $log_{10}(y)\sim-7$ to $-6.5$ - this is due to the high-density peaks not being fully
ionized in this scenario. Patchy reionization proceeds in an inside-out fashion
\citep{iliev:2006a}, with the denser regions ionized earlier, on average, compared to
the low-density ones.

In summary, compared to the base density-fluctuations-only scenario, the ionization
fraction fluctuations are responsible for the broad peak of the PDF distribution, while
eliminating the modest high-y tail due to the density peaks being ionized first, while
the temperature fluctuations yield the high-y tail in the distribution, with some values
reaching as high as $log_{10}(y)\sim-6$ to $-5.5$.



\subsection{Phase diagrams}
\label{sec:2D_results}

\begin{figure*}
    \centering
      \includegraphics[width=0.49\textwidth]{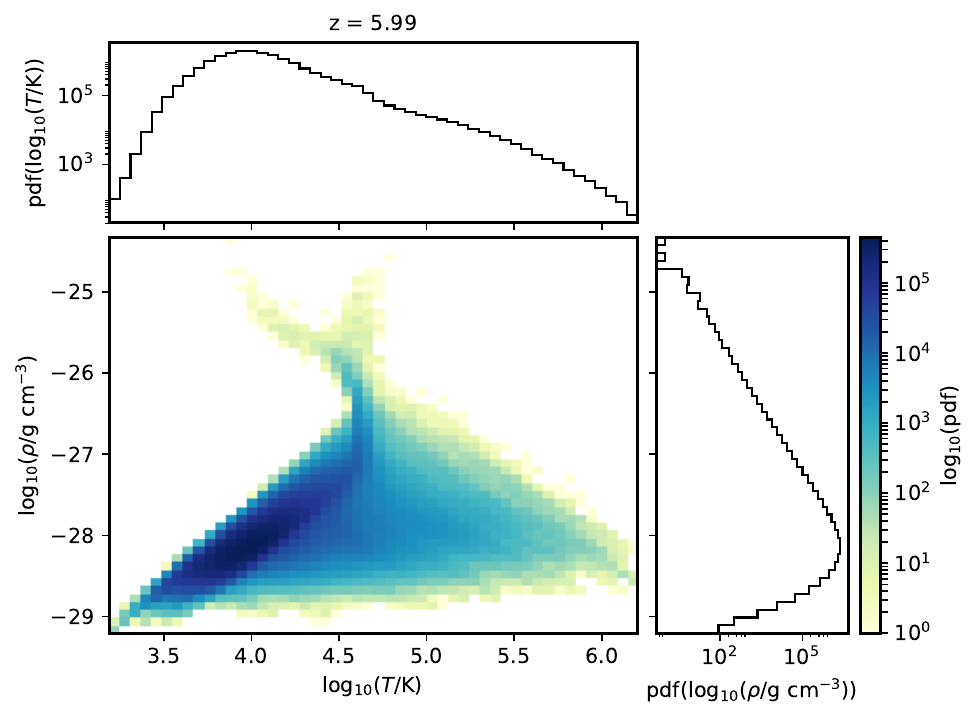}
        \label{fig:2D_simB}         
        \includegraphics[width=0.49\textwidth]{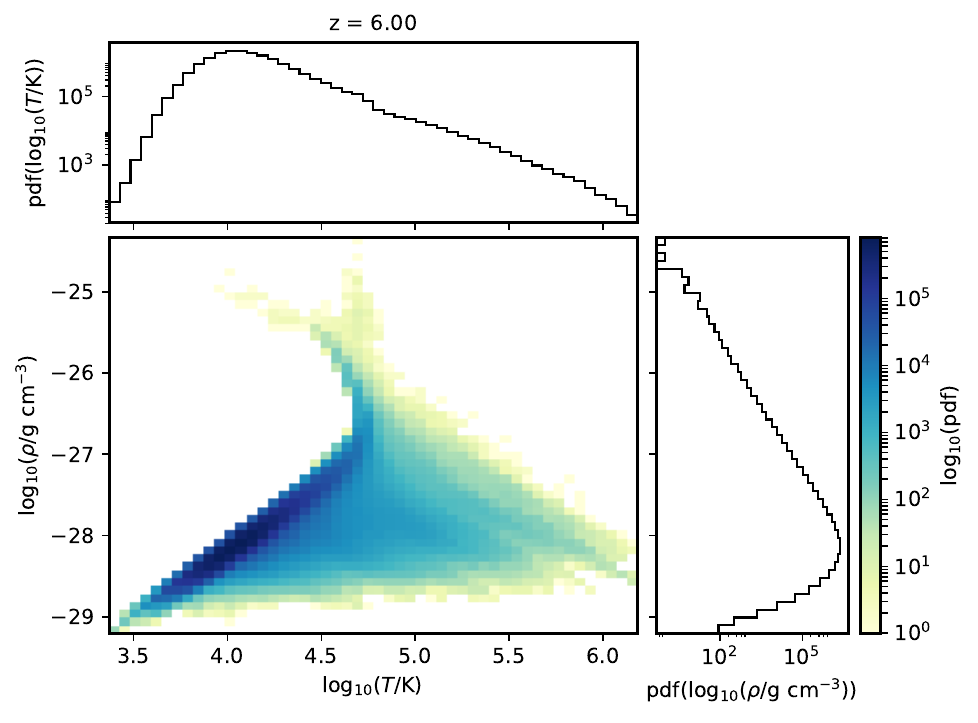}
        \label{fig:2D_simC}         
%
     \includegraphics[width=0.49\textwidth]{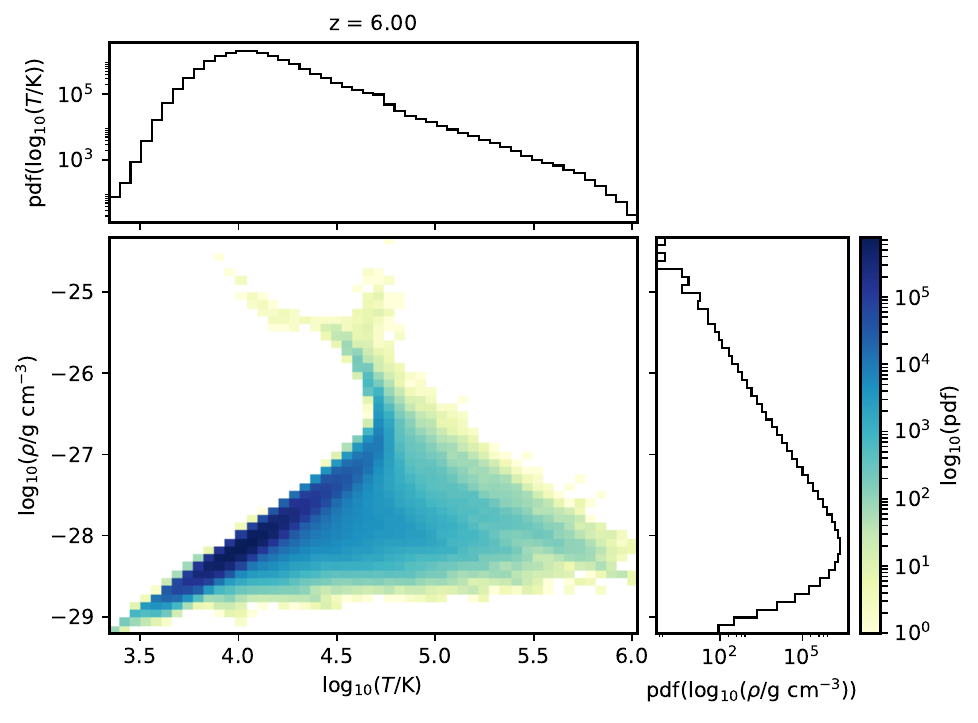}
        \label{fig:2D_simG} 
       \includegraphics[width=0.49\textwidth]{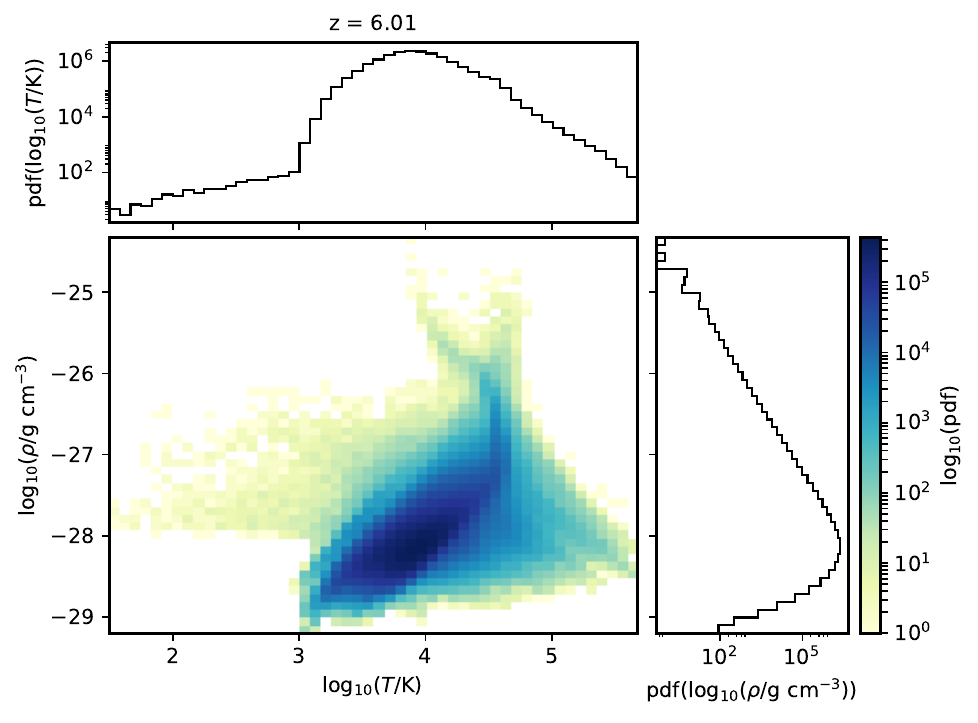}
        \label{fig:2D_simH}
        \caption{Phase diagrams showing the gas density vs. temperature at the end of the
          reionization ($z \sim 6$), for selected subset of our simulations with same resolution,
          but varying parameters:
          (top left) \simB, (top right) \simC,
          (bottom left) \simG, and (bottom right) \simH.}
    \label{fig:phase_diagrams3}
\end{figure*}

In this section, we present phase diagrams of the gas temperature vs. density (equation of state)
for a selected sub-set of our simulations (Figs.~\ref{fig:phase_diagrams1} and \ref{fig:phase_diagrams3}).
These plots reflect the local state of the gas and ultimately determine the range and variety of
y-parameters we observed. The distributions are considered at $z\sim6$, i.e. around, or soon after,
overlap. 

There are a number of features common for all cases, along with some instructive differences.
The majority of cells lie along a diagonal line from $\log_{10}T\sim3.5$ to $4.5$, corresponding
to photoionization equlibrium in the under-dense regions ($\log_{10}\rho<-26.5\,cm^{-3}$), where
gas atomic line cooling is not very strong. Volumes with higher than average density have shorter
cooling times, but are also continuously heated by stellar radiation, keeping the temperature
to a roughly constant value, independent of the local density, but varying between simulations.
We note that the observed bifurcation of the distribution at high densities in some of the cases
is an (unphysical) resolution effect, which largely disappears as the resolution increases --
comparing \hunC \,\,  vs. \fifty \,\,vs. the high-resolution runs CoDa~II and \tenA.

A relatively small fraction of the cells have very high temperatures ($T>10^5\,$K), which cannot
be reached through photoionization and are instead due to local heating from SNe explosions and
strong structure formation shocks. Although rare, these regions are important since they are the
origin of the very high y-parameter values we find along certain lines of sight. Such regions
are present in all cases, however the detailed distribution depends on the volume, resolution and
parameters of the simulation. Some cells in the higher-resolution simulations reach significantly
higher temperatures, up to $T>10^6~\,K$ or even $10^7~$K, which is not the case for the lower-resolution
cases, where the energy input is averaged over larger, coarser cells. For the same reason the later
cases also lack high-density, high-temperature cells, which are the ones that have the highest 
gas pressure and thus yield largest y-parameter values. This observation explains the high-y
tail in the y-parameter PDF distribution for CoDa-II. The small, high-resolution volumes, which
are better resolved than CoDa~II, do not reach quite such high y-values due to their much smaller
volumes, which limits the statistical sampling of different environments.

Turning our attention to the further cases in Fig.~\ref{fig:phase_diagrams3}, these provide
additional insight on how the variations in simulation input parameters affect these phase
distributions. Here we focus on comparing simulations at fixed resolution, as opposed to
the effects of varying the numerical resolution. which we have already discussed above.

The first three cases (\simB, \simC, \simG) correspond to early reionization, while the last
one (\simH) is a late-reionization one. The main variation among the former cases is in the
thickness of the photoionization equilibrium 'line' discussed above. That feature is considerably
thicker for \simB, the reason being that it completed reionization later ($z\sim6.2$), than
\simC and \simG \, ($z\sim6.6$) and thus in the latter cases by $z\sim6$ many more cells have
had time to cool down to lower equlibrium temperatures than those reached immediately post-ionization.
Other, minor differences are observed in the maximum temperatures reached ($T>10^6\,$K) in the first
two cases vs. $T<10^6\,$K for \simG, related to the lower star formation efficiency in the latter case.
Finally, in the late reionization case, \simH\,, there is a (weak) tail of low-temperature, still
neutral cells due to reionization just completing at that time. We also observe similar trends as
above, with a thicker equilibrium 'line' due to incomplete equilibration and relatively lower
maximum temperatures due to the even lower star formation efficiency in this case.

\subsection{Second-order Doppler contributions}
\begin{figure}
    \centering
        \includegraphics[width=\columnwidth]{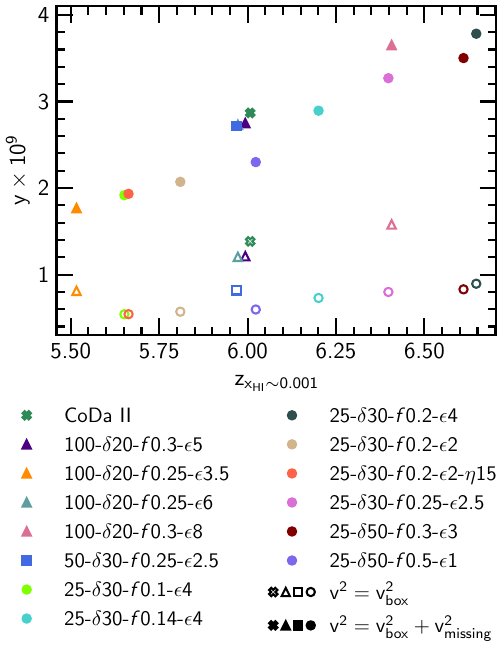}
    \caption{As in Fig.~\ref{fig:pdf_y_z}, but for the contribution to the $y$-parameter from second-order Doppler distortions. Unfilled markers correspond to the $y$-parameter computed using only the velocity of the box, while filled markers include an extra contribution to the velocity to account for missing large-scale power due to small box sizes. Note that the $100$-$\delta20$-$f0.25$-$\epsilon6$ and $50$-$\delta30$-$f0.25$-$\epsilon2.5$ cases fall directly on top of one another when including the missing power.}
    \label{fig:pdf_y_z_so}
\end{figure}

In Fig.~\ref{fig:pdf_y_z_so} we show the additional contribution to the $y$-parameter from the quadratic Doppler distortions, both as computed directly from our simulations (unfilled markers) and when accounting for the additional velocity power missing in our simulation volumes (filled markers). In both cases  there is a clear positive correlation between the redshift of the end of reionistion $z_{x_{\rm HI} \sim 0.001}$ and $y$ from quadratic Doppler. There is a notable scatter around similar $z_{x_{\rm HI} \sim 0.001}$ between the different simulation volumes. However, after accounting for the missing velocity power, this scatter is reduced significantly,  indicating that it is largely due to the different box sizes used.

The relationship between this additional Compton-$y$ contribution and the redhift of the end of reionisation  can be easily understood through equation~\ref{eq:ysoz}, where the only terms that vary as a function of $z$ between simulations are $\langle v^2 \rangle$ and $n_{\rm e}$. After accounting for missing power, the  behaviour of $\langle v^2 \rangle$ will be largely similar between all simulations, and so  the key determining factor is the evolution of $n_{\rm e}$ -- simulations that reionise earlier will have larger values of $n_{\rm e}$ throughout cosmic time and thus will have a larger $y$.  The redshift at which reionisation ends does not tell the whole story, however. Simulations can complete reionisation at the same time but have a different evolution of $n_{\rm e}$, leading to the remaining scatter we see in the $y$-parameter for simulations with similar $z_{x_{\rm HI} \sim 0.001}$. For example, from Fig.~\ref{fig:calib_plots} we see that the $100$-$\delta20$-$f0.25$-$\epsilon6$ and $50$-$\delta30$-$f0.25$-$\epsilon2.5$ simulations have very similar reionisation histories, and end up with very similar $y$-parameters (after accounting for missing velocity power). In contrast to these two simulations, reionisation begins more gradually in the $25$-$\delta50$-$f0.5$-$\epsilon1$ simulation before ending at a similar time to both $100$-$\delta20$-$f0.25$-$\epsilon6$ and $50$-$\delta30$-$f0.25$-$\epsilon2.5$. Consequently, the $y$-parameter is smaller in $25$-$\delta50$-$f0.5$-$\epsilon1$ than in $100$-$\delta20$-$f0.25$-$\epsilon6$ and $50$-$\delta30$-$f0.25$-$\epsilon2.5$.

In summary, we find that, after accounting for missing velocity power, the Compton $y$-parameter from the quadratic Doppler distortions is of order 10\% of the corresponding EoR tSZ signal, with the additional contribution ranging from $1.8 < y/10^{-9} < 3.8$, where the exact value depends on the detailed reionisation history. This means the second-order contribution is a sub-dominant, but non-trivial contribution to the EoR tSZ signal.

\section{Summary and Conclusions}\label{sec:conculsion}

The tSZ effect traces the integrated gas pressure history of the universe. The full tSZ signal is generally dominated by the contribution of galaxy clusters at lower redshifts, however we have shown here
that it also receives a notable addition from the dense, ionised IGM during the EoR. Consequently,
we argued that it is essential to investigate the extent of this effect at this stage in cosmic
history. A proper, in-depth understanding of the EoR contribution enables us to better model
the total tSZ signal and to interpret it correctly. Since the tSZ power spectrum is sensitive
to the underlying cosmological model, advancements in our measurements and models will allow for
tighter constraints to be placed on cosmological parameters, e.g. $\sigma_8$. In this work, we
utilised the data from sixteen RAMSES-CUDATON full radiative hydrodynamics cosmological simulations to compute the tSZ signal arising from the EoR. All simulations were well calibrated to yield plausible reionization histories which agree with the available observational constraints. We constructed lightcones of the electron pressure in the redshift range of reionization (roughly $z \sim 6 - 12$) by interpolating ionised fraction and gas pressure fields between simulation snapshots. From these lightcones, we calculated the Compton y-parameter by numerically integrating along lines of sight in the direction of light propagation. We obtained both high-resolution and SPT beam-smoothed maps of the y-parameter in multiple frequency channels, which we then used to construct PDFs and angular power spectra. We also estimated the contribution to the $y$-parameter due to quadratic Doppler distortions, correcting for missing velocity power due to small box sizes.

For further understanding and additional tests of the data we evaluated the contributions to the Compton y-parameter from different stages of EoR. We also probed the extent to which the temperature and density fluctuations, and patchiness of the EoR were each responsible for the values found, and considered in detail the gas equation of state phase diagrams. 

All simulations yielded mean Compton y-parameter values in the order of
$\langle y\rangle \sim \textrm{few} \times 10^{-8}$,
in rough agreement with previous estimate made by \citet{Hill:2015}. 
The magnitude of the tSZ signal originating from the EoR shows clear dependence, roughly
proportional, to the redshift of overlap/end of reionization, $z_{0.001}$ (Fig.~\ref{fig:pdf_y_z}).
Simulations which reionise early yield higher y-parameters, by up to factor of 2-3, than those where EoR ended later. For example, when the EoR ended at $z_{ov}\sim6.6$, y-parameter is up to $y\sim3.5\times10^{-8}$, while for $z_{ov}\sim5.6$, the value of $y$ can be as low as
$y\sim 1.22\times 10^{-8}$ . We also note that other parameters which influence the detailed reionization history also influence the tSZ results, increasing the scatter in the mean $y$-parameter values at a given $z_{ov}$. For example, higher star formation efficiency yields a somewhat stronger tSZ effect, likely due to increased supernovae rate. We find that the contribution of quadratic Doppler distortions to the $y$-parameter is sub-dominant to the electron pressure contribution, at the order of $\sim {\rm few} \times 10^9$, or an additional $\sim10~{\rm per~cent}$ contribution to the EoR tSZ effect. The variation in this quadratic Doppler $y$-parameter is driven by reionisation history, where scenarios that reionise earlier have a larger $y$.

For the purpose of consistency, we attempted to use the same redshift range for the lightcones. However, we note that in doing so, we did not capture the full extent of the EoR for some of our simulations. As we saw from separating the redshift contributions for the CoDa~II simulation, the majority of the tSZ signal arises around the late stages of the EoR. Another limiting factor in this study is the shorter lightcones of the 10~Mpc boxes since we were not able to run one of these simulations to the complete end of the EoR, thus somewhat underestimating the signal in that case.

The tSZ signal from EoR differs significantly from the post-reionization one in both
spatial scales and magnitude, as evidenced by the maps and power spectra of the signal that we produced. The mean signal is of order up to few percent of the cluster one, but its power spectrum
shape is quite different, with peak power at smaller scales ($\ell\sim10^5$ as opposed to
$\ell\sim\textrm{few}\times10^3$ for post-EoR signal. Overall, the EoR tSZ is a modest
contaminant at larger scales, but potentially a dominant one at smaller ones. The
Compton y-parameter maps smoothed with Gaussian beams of FWHM 1.2 arcmin and 1.7 arcmin
(corresponding to SPT bands) peak at $y\sim \textrm{few}\times10^{-8}$, roughly three orders
of magnitude lower than the typical cluster values. At these scales, the EoR contribution to
the angular power spectrum of the tSZ signal is below 1\% that of clusters. However, for
the large-volume, high-resolution CoDa~II simulation the power spectrum increases by about
one order of magnitude at smaller scales, while the post-EoR signal decreases significantly,
suggesting a more significant and potentially dominant EoR contribution there, although we note that the post-reionization tSZ is not yet fully understood at those scales.
The EoR tSZ signal peaks at such small scales in significant part due to strong local heating
by supernovae. Based on the phase diagrams and our results from the density lightcones, we see
that the high cell temperatures in dense cells reached by CoDa~II are responsible for its locally
higher y-parameter values. This tail in its PDF is a result of the spatially resolved cells heated
by supernova explosions. Removing the variations in the temperature and ionisation
fraction fields raises the mean Compton parameter. This is because they widen the range
of y-parameter values in the signal. The temperature fluctuations are responsible for
the distributions being positively skewed, due to supernova explosions driving up the
temperatures of small areas. In conclusion, while we find that the EoR contribution is
generally sub-dominant to that of galaxy clusters, it is still essential to obtain an estimate
of it as more sensitive technologies are developed in the advancing era of precision cosmology.


\section*{Acknowledgements}


ITI was supported by the Science and Technology Facilities Council [grant numbers  ST/T000473/1, ST/F002858/1 and ST/I000976/1] and the Southeast Physics Network (SEPNet). LC is supported by 
the Science and Technology Facilities Council [grant number ST/X000982/1]. KA is supported by NRF-2021R1A2C1095136 and RS-2022-00197685.
Some of the analysis was done on the Apollo2 cluster at The University of Sussex. The authors gratefully acknowledge the Gauss Centre for Supercomputing e.V. (www.gauss-centre.eu) for funding this project by providing computing time through the John von Neumann Institute for Computing (NIC) on the GCS Supercomputer JUWELS at J\"ulich Supercomputing Centre (JSC). We also acknowledge support from Partnership for Advanced Computing in Europe (PRACE) though the DECI-14 funded project subgridEoR. 

\section*{Data Availability}
The data used in this paper is available for any reasonable requests to the
corresponding author.



\bibliographystyle{mnras}
\bibliography{refs} 

\begin{thebibliography}{}
\makeatletter
\relax
\def\mn@urlcharsother{\let\do\@makeother \do\$\do\&\do\#\do\^\do\_\do\%\do\~}
\def\mn@doi{\begingroup\mn@urlcharsother \@ifnextchar [ {\mn@doi@}
  {\mn@doi@[]}}
\def\mn@doi@[#1]#2{\def\@tempa{#1}\ifx\@tempa\@empty \href
  {http://dx.doi.org/#2} {doi:#2}\else \href {http://dx.doi.org/#2} {#1}\fi
  \endgroup}
\def\mn@eprint#1#2{\mn@eprint@#1:#2::\@nil}
\def\mn@eprint@arXiv#1{\href {http://arxiv.org/abs/#1} {{\tt arXiv:#1}}}
\def\mn@eprint@dblp#1{\href {http://dblp.uni-trier.de/rec/bibtex/#1.xml}
  {dblp:#1}}
\def\mn@eprint@#1:#2:#3:#4\@nil{\def\@tempa {#1}\def\@tempb {#2}\def\@tempc
  {#3}\ifx \@tempc \@empty \let \@tempc \@tempb \let \@tempb \@tempa \fi \ifx
  \@tempb \@empty \def\@tempb {arXiv}\fi \@ifundefined
  {mn@eprint@\@tempb}{\@tempb:\@tempc}{\expandafter \expandafter \csname
  mn@eprint@\@tempb\endcsname \expandafter{\@tempc}}}

\bibitem[\protect\citeauthoryear{{Arnaud}, {Pointecouteau}  \&
  {Pratt}}{{Arnaud} et~al.}{2005}]{Arnaud:2005}
{Arnaud} M.,  {Pointecouteau} E.,   {Pratt} G.~W.,  2005, \mn@doi [\aap]
  {10.1051/0004-6361:20052856}, \href
  {https://ui.adsabs.harvard.edu/abs/2005A%26A...441..893A} {441, 893}

\bibitem[\protect\citeauthoryear{{Aubert} \& {Teyssier}}{{Aubert} \&
  {Teyssier}}{2008}]{Aubert:2008}
{Aubert} D.,  {Teyssier} R.,  2008, \mn@doi [\mnras]
  {10.1111/j.1365-2966.2008.13223.x}, \href
  {http://adsabs.harvard.edu/abs/2008MNRAS.387..295A} {387, 295}

\bibitem[\protect\citeauthoryear{{Barbosa}, {Bartlett}, {Blanchard}  \&
  {Oukbir}}{{Barbosa} et~al.}{1996}]{Barbosa:1996}
{Barbosa} D.,  {Bartlett} J.~G.,  {Blanchard} A.,   {Oukbir} J.,  1996, \aap,
  \href {https://ui.adsabs.harvard.edu/abs/1996A%26A...314...13B} {314, 13}

\bibitem[\protect\citeauthoryear{{Barkana} \& {Loeb}}{{Barkana} \&
  {Loeb}}{2001}]{Barkana2001}
{Barkana} R.,  {Loeb} A.,  2001, \mn@doi [\physrep]
  {10.1016/S0370-1573(01)00019-9}, \href
  {https://ui.adsabs.harvard.edu/abs/2001PhR...349..125B} {349, 125}

\bibitem[\protect\citeauthoryear{{Battaglia}, {Bond}, {Pfrommer}, {Sievers}  \&
  {Sijacki}}{{Battaglia} et~al.}{2010}]{Battaglia:2010}
{Battaglia} N.,  {Bond} J.~R.,  {Pfrommer} C.,  {Sievers} J.~L.,   {Sijacki}
  D.,  2010, \mn@doi [\apj] {10.1088/0004-637X/725/1/91}, \href
  {http://adsabs.harvard.edu/abs/2010ApJ...725...91B} {725, 91}

\bibitem[\protect\citeauthoryear{{Battaglia}, {Trac}, {Cen}  \&
  {Loeb}}{{Battaglia} et~al.}{2013}]{Battaglia:2013}
{Battaglia} N.,  {Trac} H.,  {Cen} R.,   {Loeb} A.,  2013, \mn@doi [\apj]
  {10.1088/0004-637X/776/2/81}, \href
  {http://adsabs.harvard.edu/abs/2013ApJ...776...81B} {776, 81}

\bibitem[\protect\citeauthoryear{Becker \& Bolton}{Becker \&
  Bolton}{2013}]{Becker_2013}
Becker G.~D.,  Bolton J.~S.,  2013, \mn@doi [Monthly Notices of the Royal
  Astronomical Society] {10.1093/mnras/stt1610}, 436, 1023

\bibitem[\protect\citeauthoryear{{Birkinshaw}}{{Birkinshaw}}{1999}]{Birkinshaw:1999}
{Birkinshaw} M.,  1999, \mn@doi [\physrep] {10.1016/S0370-1573(98)00080-5},
  \href {http://adsabs.harvard.edu/abs/1999PhR...310...97B} {310, 97}

\bibitem[\protect\citeauthoryear{{Bolliet}, {Comis}, {Komatsu}  \&
  {Mac{\'\i}as-P{\'e}rez}}{{Bolliet} et~al.}{2018}]{Bolliet:2018}
{Bolliet} B.,  {Comis} B.,  {Komatsu} E.,   {Mac{\'\i}as-P{\'e}rez} J.~F.,
  2018, \mn@doi [\mnras] {10.1093/mnras/sty823}, \href
  {https://ui.adsabs.harvard.edu/abs/2018MNRAS.477.4957B} {477, 4957}

\bibitem[\protect\citeauthoryear{Bouwens et~al.,}{Bouwens
  et~al.}{2015}]{Bouwens:2015}
Bouwens R.~J.,  et~al., 2015, \mn@doi [The Astrophysical Journal]
  {10.1088/0004-637x/803/1/34}, 803, 34

\bibitem[\protect\citeauthoryear{Calverley, Becker, Haehnelt  \&
  Bolton}{Calverley et~al.}{2011}]{Calverley:2011}
Calverley A.~P.,  Becker G.~D.,  Haehnelt M.~G.,   Bolton J.~S.,  2011, \mn@doi
  [\mnras] {10.1111/j.1365-2966.2010.18072.x}, 412, 2543

\bibitem[\protect\citeauthoryear{{Carlstrom}, {Holder}  \& {Reese}}{{Carlstrom}
  et~al.}{2002}]{Carlstrom:2002}
{Carlstrom} J.~E.,  {Holder} G.~P.,   {Reese} E.~D.,  2002, \mn@doi [\araa]
  {10.1146/annurev.astro.40.060401.093803}, \href
  {http://adsabs.harvard.edu/abs/2002ARA\%26A..40..643C} {40, 643}

\bibitem[\protect\citeauthoryear{{Carlstrom} et~al.,}{{Carlstrom}
  et~al.}{2011}]{Carlstrom:2011}
{Carlstrom} J.~E.,  et~al., 2011, \mn@doi [\pasp] {10.1086/659879}, \href
  {https://ui.adsabs.harvard.edu/abs/2011PASP..123..568C} {123, 568}

\bibitem[\protect\citeauthoryear{{Challinor} \& {Lasenby}}{{Challinor} \&
  {Lasenby}}{1998}]{Challinor:1998}
{Challinor} A.,  {Lasenby} A.,  1998, \mn@doi [\apj] {10.1086/305623}, \href
  {https://ui.adsabs.harvard.edu/abs/1998ApJ...499....1C} {499, 1}

\bibitem[\protect\citeauthoryear{Chluba, Switzer, Nelson  \& Nagai}{Chluba
  et~al.}{2013}]{Chluba:2013}
Chluba J.,  Switzer E.,  Nelson K.,   Nagai D.,  2013, \mn@doi [\mnras]
  {10.1093/mnras/stt110}, 430, 3054

\bibitem[\protect\citeauthoryear{{Crawford} et~al.,}{{Crawford}
  et~al.}{2014}]{Crawford:2014}
{Crawford} T.~M.,  et~al., 2014, \mn@doi [\apj] {10.1088/0004-637X/784/2/143},
  \href {https://ui.adsabs.harvard.edu/abs/2014ApJ...784..143C} {784, 143}

\bibitem[\protect\citeauthoryear{{D'Aloisio}, {McQuinn}, {Maupin}, {Davies},
  {Trac}, {Fuller}  \& {Upton Sanderbeck}}{{D'Aloisio}
  et~al.}{2019}]{DAloisio:2019}
{D'Aloisio} A.,  {McQuinn} M.,  {Maupin} O.,  {Davies} F.~B.,  {Trac} H.,
  {Fuller} S.,   {Upton Sanderbeck} P.~R.,  2019, \mn@doi [\apj]
  {10.3847/1538-4357/ab0d83}, \href
  {https://ui.adsabs.harvard.edu/abs/2019ApJ...874..154D} {874, 154}

\bibitem[\protect\citeauthoryear{Davies et~al.,}{Davies
  et~al.}{2018}]{Davies_2018}
Davies F.~B.,  et~al., 2018, \mn@doi [The Astrophysical Journal]
  {10.3847/1538-4357/aad6dc}, 864, 142

\bibitem[\protect\citeauthoryear{{Eisenstein} \& {Hu}}{{Eisenstein} \&
  {Hu}}{1999}]{1999ApJ...511....5E}
{Eisenstein} D.~J.,  {Hu} W.,  1999, \mn@doi [\apj] {10.1086/306640}, \href
  {https://ui.adsabs.harvard.edu/abs/1999ApJ...511....5E} {511, 5}

\bibitem[\protect\citeauthoryear{{Fan}, {Carilli}  \& {Keating}}{{Fan}
  et~al.}{2006}]{Fan:2006}
{Fan} X.,  {Carilli} C.~L.,   {Keating} B.,  2006, \mn@doi [\araa]
  {10.1146/annurev.astro.44.051905.092514}, \href
  {https://ui.adsabs.harvard.edu/abs/2006ARA%26A..44..415F} {44, 415}

\bibitem[\protect\citeauthoryear{Faucher-Gigu\'ere, Lidz, Hernquist  \&
  Zaldarriaga}{Faucher-Gigu\'ere et~al.}{2008}]{Faucher_Giguere_2008}
Faucher-Gigu\'ere C.,  Lidz A.,  Hernquist L.,   Zaldarriaga M.,  2008, \mn@doi
  [The Astrophysical Journal] {10.1086/592289}, 688, 85

\bibitem[\protect\citeauthoryear{{Fixsen}, {Cheng}, {Gales}, {Mather}, {Shafer}
   \& {Wright}}{{Fixsen} et~al.}{1996}]{Fixsen:1996}
{Fixsen} D.~J.,  {Cheng} E.~S.,  {Gales} J.~M.,  {Mather} J.~C.,  {Shafer}
  R.~A.,   {Wright} E.~L.,  1996, \mn@doi [\apj] {10.1086/178173}, \href
  {https://ui.adsabs.harvard.edu/abs/1996ApJ...473..576F} {473, 576}

\bibitem[\protect\citeauthoryear{{George} et~al.,}{{George}
  et~al.}{2015}]{George:2015}
{George} E.~M.,  et~al., 2015, \mn@doi [\apj] {10.1088/0004-637X/799/2/177},
  \href {http://adsabs.harvard.edu/abs/2015ApJ...799..177G} {799, 177}

\bibitem[\protect\citeauthoryear{Greig \& Mesinger}{Greig \&
  Mesinger}{2017}]{Greig_2017}
Greig B.,  Mesinger A.,  2017, \mn@doi [Monthly Notices of the Royal
  Astronomical Society] {10.1093/mnras/stw3026}, 465, 4838

\bibitem[\protect\citeauthoryear{{Hill} et~al.,}{{Hill}
  et~al.}{2014}]{Hill:2014}
{Hill} J.~C.,  et~al., 2014, arXiv e-prints, ArXiv/1411.8004, \href
  {https://ui.adsabs.harvard.edu/abs/2014arXiv1411.8004H} {}

\bibitem[\protect\citeauthoryear{{Hill}, {Battaglia}, {Chluba}, {Ferraro},
  {Schaan}  \& {Spergel}}{{Hill} et~al.}{2015}]{Hill:2015}
{Hill} J.~C.,  {Battaglia} N.,  {Chluba} J.,  {Ferraro} S.,  {Schaan} E.,
  {Spergel} D.~N.,  2015, \mn@doi [Physical Review Letters]
  {10.1103/PhysRevLett.115.261301}, \href
  {http://adsabs.harvard.edu/abs/2015PhRvL.115z1301H} {115, 261301}

\bibitem[\protect\citeauthoryear{Hoag et~al.,}{Hoag et~al.}{2019}]{Hoag_2019}
Hoag A.,  et~al., 2019, \mn@doi [The Astrophysical Journal]
  {10.3847/1538-4357/ab1de7}, 878, 12

\bibitem[\protect\citeauthoryear{{Hogg}}{{Hogg}}{1999}]{Hogg:1999}
{Hogg} D.~W.,  1999, arXiv Astrophysics e-prints, \href
  {http://adsabs.harvard.edu/abs/1999astro.ph..5116H} {}

\bibitem[\protect\citeauthoryear{Horowitz \& Seljak}{Horowitz \&
  Seljak}{2017}]{Horowitz:2017}
Horowitz B.,  Seljak U.,  2017, \mn@doi [\mnras] {10.1093/mnras/stx766}, 469,
  394

\bibitem[\protect\citeauthoryear{{Hu}, {Scott}  \& {Silk}}{{Hu}
  et~al.}{1994}]{Hu:1994a}
{Hu} W.,  {Scott} D.,   {Silk} J.,  1994, \mn@doi [\prd]
  {10.1103/PhysRevD.49.648}, \href
  {http://adsabs.harvard.edu/abs/1994PhRvD..49..648H} {49, 648}

\bibitem[\protect\citeauthoryear{{Iliev}, {Mellema}, {Pen}, {Merz}, {Shapiro}
  \& {Alvarez}}{{Iliev} et~al.}{2006a}]{Iliev:2006}
{Iliev} I.~T.,  {Mellema} G.,  {Pen} U.-L.,  {Merz} H.,  {Shapiro} P.~R.,
  {Alvarez} M.~A.,  2006a, \mn@doi [\mnras] {10.1111/j.1365-2966.2006.10502.x},
  \href {https://ui.adsabs.harvard.edu/abs/2006MNRAS.369.1625I} {369, 1625}

\bibitem[\protect\citeauthoryear{{Iliev}, {Mellema}, {Pen}, {Merz}, {Shapiro}
  \& {Alvarez}}{{Iliev} et~al.}{2006b}]{iliev:2006a}
{Iliev} I.~T.,  {Mellema} G.,  {Pen} U.-L.,  {Merz} H.,  {Shapiro} P.~R.,
  {Alvarez} M.~A.,  2006b, \mn@doi [\mnras] {10.1111/j.1365-2966.2006.10502.x},
  \href {https://ui.adsabs.harvard.edu/abs/2006MNRAS.369.1625I} {369, 1625}

\bibitem[\protect\citeauthoryear{{Iliev}, {Pen}, {Bond}, {Mellema}  \&
  {Shapiro}}{{Iliev} et~al.}{2007}]{Iliev:2007}
{Iliev} I.~T.,  {Pen} U.-L.,  {Bond} J.~R.,  {Mellema} G.,   {Shapiro} P.~R.,
  2007, \mn@doi [\apj] {10.1086/513687}, \href
  {http://adsabs.harvard.edu/abs/2007ApJ...660..933I} {660, 933}

\bibitem[\protect\citeauthoryear{{Iliev}, {Mellema}, {Pen}, {Bond}  \&
  {Shapiro}}{{Iliev} et~al.}{2008}]{Iliev:2008}
{Iliev} I.~T.,  {Mellema} G.,  {Pen} U.-L.,  {Bond} J.~R.,   {Shapiro} P.~R.,
  2008, \mn@doi [\mnras] {10.1111/j.1365-2966.2007.12629.x}, \href
  {http://adsabs.harvard.edu/abs/2008MNRAS.384..863I} {384, 863}

\bibitem[\protect\citeauthoryear{{Jeli{\'c}} et~al.,}{{Jeli{\'c}}
  et~al.}{2010}]{Jelic2010}
{Jeli{\'c}} V.,  et~al., 2010, \mn@doi [\mnras]
  {10.1111/j.1365-2966.2009.16086.x}, \href
  {https://ui.adsabs.harvard.edu/abs/2010MNRAS.402.2279J} {402, 2279}

\bibitem[\protect\citeauthoryear{{Kitayama}}{{Kitayama}}{2014}]{Kitayama:2014}
{Kitayama} T.,  2014, \mn@doi [Progress of Theoretical and Experimental
  Physics] {10.1093/ptep/ptu055}, \href
  {https://ui.adsabs.harvard.edu/abs/2014PTEP.2014fB111K} {2014, 06B111}

\bibitem[\protect\citeauthoryear{{Komatsu} \& {Seljak}}{{Komatsu} \&
  {Seljak}}{2002}]{Komatsu:2002}
{Komatsu} E.,  {Seljak} U.,  2002, \mn@doi [\mnras]
  {10.1046/j.1365-8711.2002.05889.x}, \href
  {http://adsabs.harvard.edu/abs/2002MNRAS.336.1256K} {336, 1256}

\bibitem[\protect\citeauthoryear{Lahav, Lilje, Primack  \& Rees}{Lahav
  et~al.}{1991}]{lahav1991}
Lahav O.,  Lilje P.~B.,  Primack J.~R.,   Rees M.~J.,  1991, \mn@doi [Monthly
  Notices of the Royal Astronomical Society] {10.1093/mnras/251.1.128}, 251,
  128

\bibitem[\protect\citeauthoryear{Mason, Treu, Dijkstra, Mesinger, Trenti,
  Pentericci, de Barros  \& Vanzella}{Mason et~al.}{2018}]{Mason_2018}
Mason C.~A.,  Treu T.,  Dijkstra M.,  Mesinger A.,  Trenti M.,  Pentericci L.,
  de Barros S.,   Vanzella E.,  2018, \mn@doi [The Astrophysical Journal]
  {10.3847/1538-4357/aab0a7}, 856, 2

\bibitem[\protect\citeauthoryear{{Mesinger}, {McQuinn}  \&
  {Spergel}}{{Mesinger} et~al.}{2012}]{Mesinger:2012}
{Mesinger} A.,  {McQuinn} M.,   {Spergel} D.~N.,  2012, \mn@doi [\mnras]
  {10.1111/j.1365-2966.2012.20713.x}, \href
  {http://adsabs.harvard.edu/abs/2012MNRAS.422.1403M} {422, 1403}

\bibitem[\protect\citeauthoryear{{Mroczkowski} et~al.,}{{Mroczkowski}
  et~al.}{2019}]{Mroczkowski:2019}
{Mroczkowski} T.,  et~al., 2019, \mn@doi [\ssr] {10.1007/s11214-019-0581-2},
  \href {https://ui.adsabs.harvard.edu/abs/2019SSRv..215...17M} {215, 17}

\bibitem[\protect\citeauthoryear{{Nozawa}, {Itoh}, {Suda}  \&
  {Ohhata}}{{Nozawa} et~al.}{2006}]{Nozawa:2006}
{Nozawa} S.,  {Itoh} N.,  {Suda} Y.,   {Ohhata} Y.,  2006, \mn@doi [Nuovo
  Cimento B Serie] {10.1393/ncb/i2005-10223-0}, \href
  {https://ui.adsabs.harvard.edu/abs/2006NCimB.121..487N} {121, 487}

\bibitem[\protect\citeauthoryear{{Ocvirk} et~al.,}{{Ocvirk}
  et~al.}{2018}]{Ocvirk:2018}
{Ocvirk} P.,  et~al., 2018, arXiv e-prints, \href
  {http://adsabs.harvard.edu/abs/2018arXiv181111192O} {}

\bibitem[\protect\citeauthoryear{{Ouchi} et~al.,}{{Ouchi}
  et~al.}{2010}]{Ouchi:2010}
{Ouchi} M.,  et~al., 2010, \mn@doi [\apj] {10.1088/0004-637X/723/1/869}, \href
  {https://ui.adsabs.harvard.edu/abs/2010ApJ...723..869O} {723, 869}

\bibitem[\protect\citeauthoryear{{Park}, {Shapiro}, {Komatsu}, {Iliev}, {Ahn}
  \& {Mellema}}{{Park} et~al.}{2013}]{Park:2013}
{Park} H.,  {Shapiro} P.~R.,  {Komatsu} E.,  {Iliev} I.~T.,  {Ahn} K.,
  {Mellema} G.,  2013, \mn@doi [\apj] {10.1088/0004-637X/769/2/93}, \href
  {http://adsabs.harvard.edu/abs/2013ApJ...769...93P} {769, 93}

\bibitem[\protect\citeauthoryear{{Planck Collaboration} et~al.,}{{Planck
  Collaboration} et~al.}{2014}]{Planck-Collaboration:2014b}
{Planck Collaboration} et~al., 2014, \mn@doi [\aap]
  {10.1051/0004-6361/201321591}, \href
  {https://ui.adsabs.harvard.edu/abs/2014A%26A...571A..16P} {571, A16}

\bibitem[\protect\citeauthoryear{{Planck Collaboration} et~al.,}{{Planck
  Collaboration} et~al.}{2015}]{Planck-Collaboration:2015b}
{Planck Collaboration} et~al., 2015, preprint, \href
  {http://adsabs.harvard.edu/abs/2015arXiv150201596P} {} (\mn@eprint {arXiv}
  {1502.01596})

\bibitem[\protect\citeauthoryear{{Planck Collaboration} et~al.,}{{Planck
  Collaboration} et~al.}{2018}]{Planck-Collaboration:2018}
{Planck Collaboration} et~al., 2018, arXiv e-prints, \href
  {http://adsabs.harvard.edu/abs/2018arXiv180706209P} {}

\bibitem[\protect\citeauthoryear{{Press} \& {Schechter}}{{Press} \&
  {Schechter}}{1974}]{Press:1974}
{Press} W.~H.,  {Schechter} P.,  1974, \mn@doi [\apj] {10.1086/152650}, \href
  {https://ui.adsabs.harvard.edu/abs/1974ApJ...187..425P} {187, 425}

\bibitem[\protect\citeauthoryear{{Refregier}, {Komatsu}, {Spergel}  \&
  {Pen}}{{Refregier} et~al.}{2000}]{Refregier:2000}
{Refregier} A.,  {Komatsu} E.,  {Spergel} D.~N.,   {Pen} U.-L.,  2000, \mn@doi
  [\prd] {10.1103/PhysRevD.61.123001}, \href
  {http://adsabs.harvard.edu/abs/2000PhRvD..61l3001R} {61, 123001}

\bibitem[\protect\citeauthoryear{{Rephaeli}}{{Rephaeli}}{1995}]{Rephaeli:1995}
{Rephaeli} Y.,  1995, \mn@doi [\araa] {10.1146/annurev.aa.33.090195.002545},
  \href {http://adsabs.harvard.edu/abs/1995ARA\%26A..33..541R} {33, 541}

\bibitem[\protect\citeauthoryear{{Rybicki} \& {Lightman}}{{Rybicki} \&
  {Lightman}}{1979}]{Rybicki:1979}
{Rybicki} G.~B.,  {Lightman} A.~P.,  1979, {Radiative processes in
  astrophysics}

\bibitem[\protect\citeauthoryear{{Sazonov} \& {Sunyaev}}{{Sazonov} \&
  {Sunyaev}}{2000}]{Sazonov:2000}
{Sazonov} S.~Y.,  {Sunyaev} R.~A.,  2000, \mn@doi [\apj] {10.1086/317078},
  \href {https://ui.adsabs.harvard.edu/abs/2000ApJ...543...28S} {543, 28}

\bibitem[\protect\citeauthoryear{{Sehgal}, {Bode}, {Das},
  {Hernandez-Monteagudo}, {Huffenberger}, {Lin}, {Ostriker}  \&
  {Trac}}{{Sehgal} et~al.}{2010}]{Sehgal:2010}
{Sehgal} N.,  {Bode} P.,  {Das} S.,  {Hernandez-Monteagudo} C.,  {Huffenberger}
  K.,  {Lin} Y.-T.,  {Ostriker} J.~P.,   {Trac} H.,  2010, \mn@doi [\apj]
  {10.1088/0004-637X/709/2/920}, \href
  {http://adsabs.harvard.edu/abs/2010ApJ...709..920S} {709, 920}

\bibitem[\protect\citeauthoryear{{Seljak}, {Burwell}  \& {Pen}}{{Seljak}
  et~al.}{2001}]{Seljak:2001}
{Seljak} U.,  {Burwell} J.,   {Pen} U.-L.,  2001, \mn@doi [\prd]
  {10.1103/PhysRevD.63.063001}, \href
  {https://ui.adsabs.harvard.edu/abs/2001PhRvD..63f3001S} {63, 063001}

\bibitem[\protect\citeauthoryear{{Shaw}, {Nagai}, {Bhattacharya}  \&
  {Lau}}{{Shaw} et~al.}{2010}]{Shaw:2010}
{Shaw} L.~D.,  {Nagai} D.,  {Bhattacharya} S.,   {Lau} E.~T.,  2010, \mn@doi
  [\apj] {10.1088/0004-637X/725/2/1452}, \href
  {http://adsabs.harvard.edu/abs/2010ApJ...725.1452S} {725, 1452}

\bibitem[\protect\citeauthoryear{{Sievers} et~al.,}{{Sievers}
  et~al.}{2013}]{Sievers:2013}
{Sievers} J.~L.,  et~al., 2013, \mn@doi [\jcap]
  {10.1088/1475-7516/2013/10/060}, \href
  {http://adsabs.harvard.edu/abs/2013JCAP...10..060S} {10, 060}

\bibitem[\protect\citeauthoryear{{Sunyaev} \& {Zeldovich}}{{Sunyaev} \&
  {Zeldovich}}{1972}]{Sunyaev:1972}
{Sunyaev} R.~A.,  {Zeldovich} Y.~B.,  1972, Comments on Astrophysics and Space
  Physics, \href {http://adsabs.harvard.edu/abs/1972CoASP...4..173S} {4, 173}

\bibitem[\protect\citeauthoryear{{Sunyaev} \& {Zeldovich}}{{Sunyaev} \&
  {Zeldovich}}{1980a}]{Sunyaev:1980a}
{Sunyaev} R.~A.,  {Zeldovich} Y.~B.,  1980a, \mn@doi [\araa]
  {10.1146/annurev.aa.18.090180.002541}, \href
  {http://adsabs.harvard.edu/abs/1980ARA\%26A..18..537S} {18, 537}

\bibitem[\protect\citeauthoryear{{Sunyaev} \& {Zeldovich}}{{Sunyaev} \&
  {Zeldovich}}{1980b}]{Sunyaev:1980b}
{Sunyaev} R.~A.,  {Zeldovich} Y.~B.,  1980b, \mn@doi [\mnras]
  {10.1093/mnras/190.3.413}, \href
  {http://adsabs.harvard.edu/abs/1980MNRAS.190..413S} {190, 413}

\bibitem[\protect\citeauthoryear{{Tashiro}, {Aghanim}, {Langer}, {Douspis},
  {Zaroubi}  \& {Jelic}}{{Tashiro} et~al.}{2010}]{Tashiro2010}
{Tashiro} H.,  {Aghanim} N.,  {Langer} M.,  {Douspis} M.,  {Zaroubi} S.,
  {Jelic} V.,  2010, \mn@doi [\mnras] {10.1111/j.1365-2966.2009.16078.x}, \href
  {https://ui.adsabs.harvard.edu/abs/2010MNRAS.402.2617T} {402, 2617}

\bibitem[\protect\citeauthoryear{{Teyssier}}{{Teyssier}}{2002}]{Teyssier:2002}
{Teyssier} R.,  2002, \mn@doi [\aap] {10.1051/0004-6361:20011817}, \href
  {http://adsabs.harvard.edu/abs/2002A\%26A...385..337T} {385, 337}

\bibitem[\protect\citeauthoryear{Toro}{Toro}{2009}]{Toro:2009}
Toro E.~F.,  2009, Riemann Solvers and Numerical Methods for Fluid Dynamics, 3
  edn.
Springer-Verlag Berlin Heidelberg, \mn@doi{10.1007/b79761}

\bibitem[\protect\citeauthoryear{{Trac}, {Bode}  \& {Ostriker}}{{Trac}
  et~al.}{2011}]{Trac:2011}
{Trac} H.,  {Bode} P.,   {Ostriker} J.~P.,  2011, \mn@doi [\apj]
  {10.1088/0004-637X/727/2/94}, \href
  {http://adsabs.harvard.edu/abs/2011ApJ...727...94T} {727, 94}

\bibitem[\protect\citeauthoryear{Wyithe \& Bolton}{Wyithe \&
  Bolton}{2011}]{Wyithe:2011}
Wyithe J. S.~B.,  Bolton J.~S.,  2011, \mn@doi [\mnras]
  {10.1111/j.1365-2966.2010.18030.x}, 412, 1926

\bibitem[\protect\citeauthoryear{{Zeldovich}, {Illarionov}  \&
  {Syunyaev}}{{Zeldovich} et~al.}{1972}]{Zeldovich:1972}
{Zeldovich} Y.~B.,  {Illarionov} A.~F.,   {Syunyaev} R.~A.,  1972, Zhurnal
  Eksperimentalnoi i Teoreticheskoi Fiziki, \href
  {https://ui.adsabs.harvard.edu/abs/1972ZhETF..62.1217Z} {62, 1217}

\bibitem[\protect\citeauthoryear{{Zhang}, {Pen}  \& {Trac}}{{Zhang}
  et~al.}{2004}]{Zhang:2004a}
{Zhang} P.,  {Pen} U.-L.,   {Trac} H.,  2004, \mn@doi [\mnras]
  {10.1111/j.1365-2966.2004.08328.x}, \href
  {http://adsabs.harvard.edu/abs/2004MNRAS.355..451Z} {355, 451}

\bibitem[\protect\citeauthoryear{{da Silva}, {Barbosa}, {Liddle}  \&
  {Thomas}}{{da Silva} et~al.}{2001}]{da-Silva:2001}
{da Silva} A.~C.,  {Barbosa} D.,  {Liddle} A.~R.,   {Thomas} P.~A.,  2001,
  \mn@doi [\mnras] {10.1046/j.1365-8711.2001.04580.x}, \href
  {http://adsabs.harvard.edu/abs/2001MNRAS.326..155D} {326, 155}

\makeatother
\end{thebibliography}






\bsp	
\end{document}